\documentclass[12pt]{iopart}

\usepackage{axodraw}
\usepackage{graphicx}
\usepackage{rotating}
\usepackage{color}

\begin{document}



\title{Searches for ttH (Multilepton+Diphoton) Production in ATLAS}

\author{Andr\'e Sopczak on behalf of the ATLAS Collaboration}

\address{Institute of Experimental and Applied Physics, Czech Technical University in Prague}
\ead{andre.sopczak@cern.ch}

\begin{center}
{\bf\small ABSTRACT}
\end{center}
{\large 
After the discovery of a Higgs boson, the measurements of its properties are 
at the forefront of research. The determination of the associated production of a 
Higgs boson and a pair of top quarks is of particular importance as the ttH Yukawa 
coupling is large and thus an excellent probe for physics beyond the Standard Model (SM). 
For the complete LHC Run-1 dataset, the combined ATLAS and CMS 
signal strength (defined as the ratio of measured cross-section to the SM prediction) was
$\mu=2.3^{+0.7}_{-0.6}$, which indicated a mild excess with respect to the SM expectation.
The results of the ttH search with multilepton and diphoton signatures are 
presented for the first LHC Run-2 data (13~fb$^{-1}$ at 13 TeV) 
recorded by the ATLAS experiment:
$\mu=2.5^{+1.3}_{-1.1}$ for the multilepton analyses and 
$\mu=-0.3^{+1.2}_{-1.0}$ for the diphoton analyses.
Both measurements are in agreement with the SM expectation.
}

\vspace*{3cm}
\begin{center}
{\em Contribution to the Annual Workshop: Higgs Coupling 2016, \\
     Menlo Park, CA, USA, 9-12 November 2016}
\end{center}

\maketitle

\setlength{\textheight}{250mm}

\newcommand{\bb}{bb}
\newcommand{\nn}{\nu\nu}
\newcommand{\tautau}{\tau^+\tau^-}
\newcommand{\ee}{\mbox{$\mathrm{e}^{+}\mathrm{e}^{-}$}}

\newcommand{\Zo} {{\mathrm {Z}}}
\newcommand{\db}    {{d_{\rm B}}}
\newcommand{\dgz}  {{\Delta g_1^{\Zo}}}
\newcommand{\dkg}   {{\Delta \kappa_\gamma}}

\newcommand{\pb}   {\mbox{$\rm pb^{-1}$}}
\newcommand{\fb}   {\mbox{$\rm fb^{-1}$}}

\clearpage
\section{Introduction}
An important motivation of the top Yukawa coupling (ttH) research is the fact that at present 
it is the only quantity which can help us to get an idea about the scale of 
New Physics~\cite{Bezrukov:2014ina}.
A fundamental prediction of the Brout-Englert-Higgs model is that the 
Yukawa coupling 
is proportional to the fermion mass $m_{\rm f}$.
As $m_{\rm t}/m_{\rm b} \approx 35$, the ttH coupling is much stronger 
than the other couplings to lighter fermions.

\begin{figure}[htb]
\vspace*{-0.2cm}
\begin{center}
\includegraphics[width=0.49\textwidth,height=7.0cm]{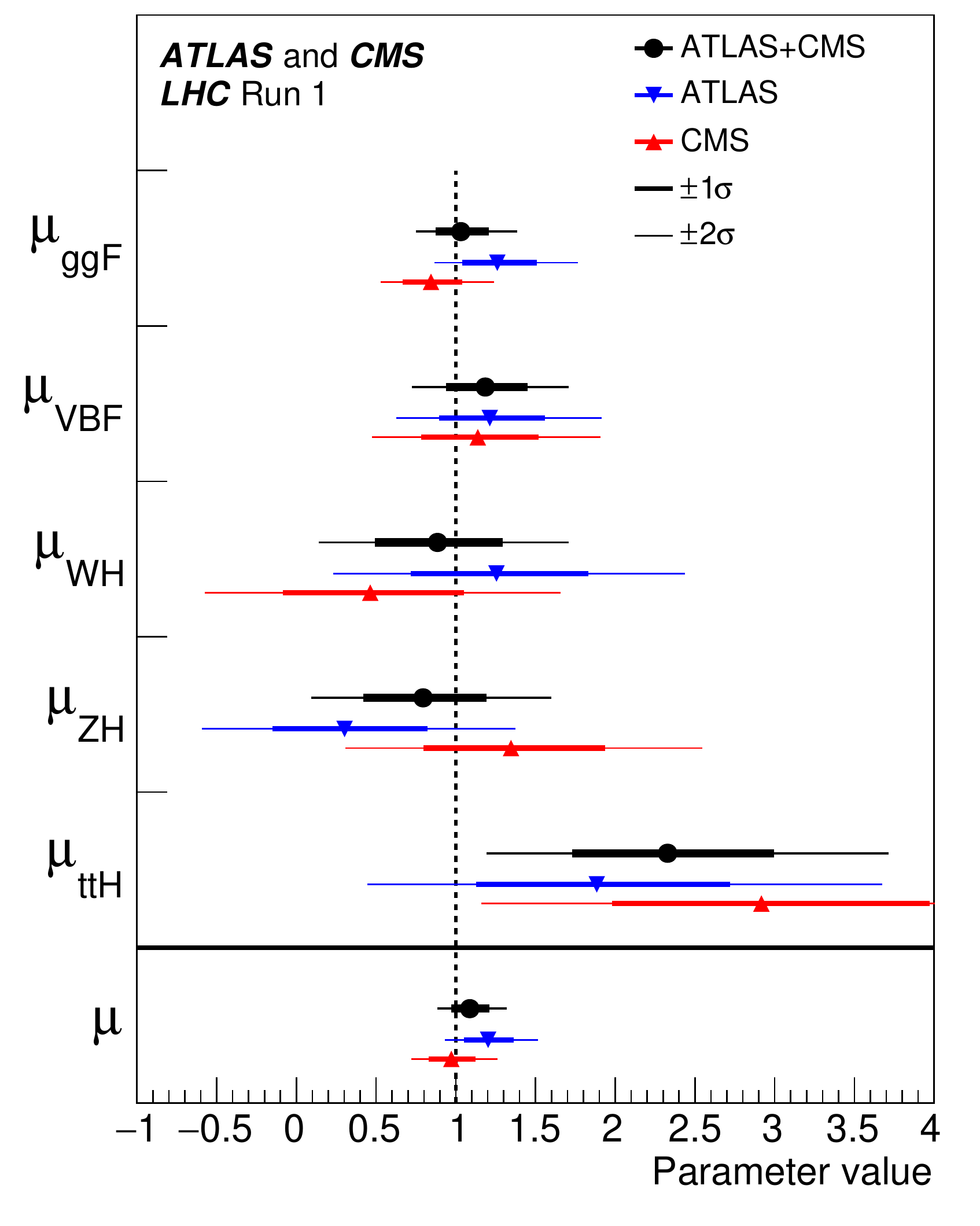} \hfill
\includegraphics[width=0.49\textwidth,height=7.3cm]{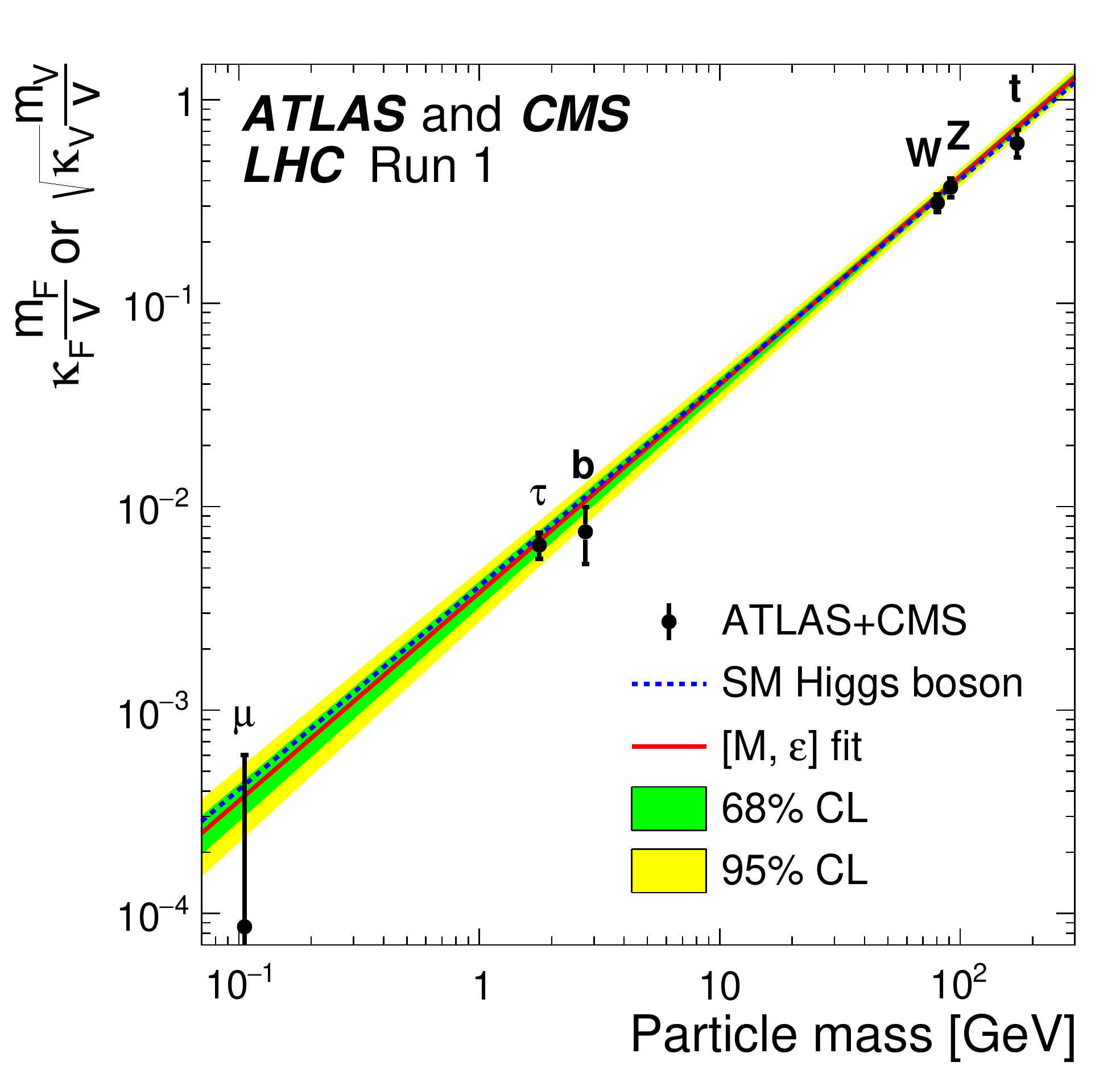}
\end{center}
\vspace*{-0.8cm}
\caption{
\newcommand{\Cc}{\ensuremath{\kappa}}
Left: best fit results for the production signal strengths for the 
combination of ATLAS and CMS data, as well as from each experiment.
The error bars indicate the $1\sigma$~(thick lines) and $2\sigma$~(thin lines) intervals.
The measurements of the global signal strength~$\mu$ are also shown.
Right: best fit values as a function of particle mass for the combination of 
ATLAS and CMS data in the case of the parameterisation described 
in~Ref.~\cite{Khachatryan:2016vau},
with parameters defined as $\Cc_{F} \cdot\ m_{F}/v$ for the fermions, 
and as $\sqrt{\Cc_{V}} \cdot\ m_{V}/v$ for the weak vector bosons, 
where $v = 246$~GeV is the vacuum expectation value of the Higgs field. 
The dashed (blue) line indicates the predicted dependence on the particle 
mass in the case of the SM~Higgs boson. The solid (red) line indicates the 
best fit result to the $[M,\epsilon]$~phenomenological model 
of~Ref.~\cite{MepsilonTheory} with the corresponding 68\%~and 95\%~CL~bands.
}
\label{fig:bm:GenMod1:prplot}
\vspace*{-0.1cm}
\end{figure}

The ATLAS and CMS collaborations~\cite{atlas,cms} have extensive programmes to search for the 
ttH signal and measure the ttH coupling strength. The results of both collaborations 
using the data of LHC Run-1 were combined and gave as ratio 
of signal strengths $\mu = \sigmaσ_{\rm ttH,obs} / \sigma_{\rm ttH,SM} = 2.3^{+0.7}_{-0.6}$.
The measurement is $2.3\sigma$ above the SM expectation.
The combined ttH signal strength is shown in
Fig.~\ref{fig:bm:GenMod1:prplot} (from~\cite{Khachatryan:2016vau})
compared to other Higgs boson signal strength measurements.
Figure~\ref{fig:bm:GenMod1:prplot} also shows the impressive confirmation within the 
uncertainties of the linearity between the Yukawa coupling and the fermion mass.
As the top Yukawa coupling appears not only in the direct measurements, but also for example
in the gluon-gluon-fusion process, the best fit of the top Yukawa coupling lies slightly below 
the diagonal line, indicating an overall smaller value than expected in the SM. 
Thus, there is some tension between the indirect (loop) ttH measurements 
and the direct (tree-level) ttH measurements. 

Regarding the ATLAS ttH multilepton searches in LHC Run-1, 
there were five channels: 
one lepton with two hadronic taus, 
two same-charge leptons with no tau, 
two same-charge leptons plus one tau, 
three leptons and four leptons which are tau inclusive. 
In the initial Run-2 data analysis, there are only the later four channels.

\begin{figure}[h!]
\vspace*{-0.4cm}
\includegraphics[width=0.49\textwidth,height=6cm]{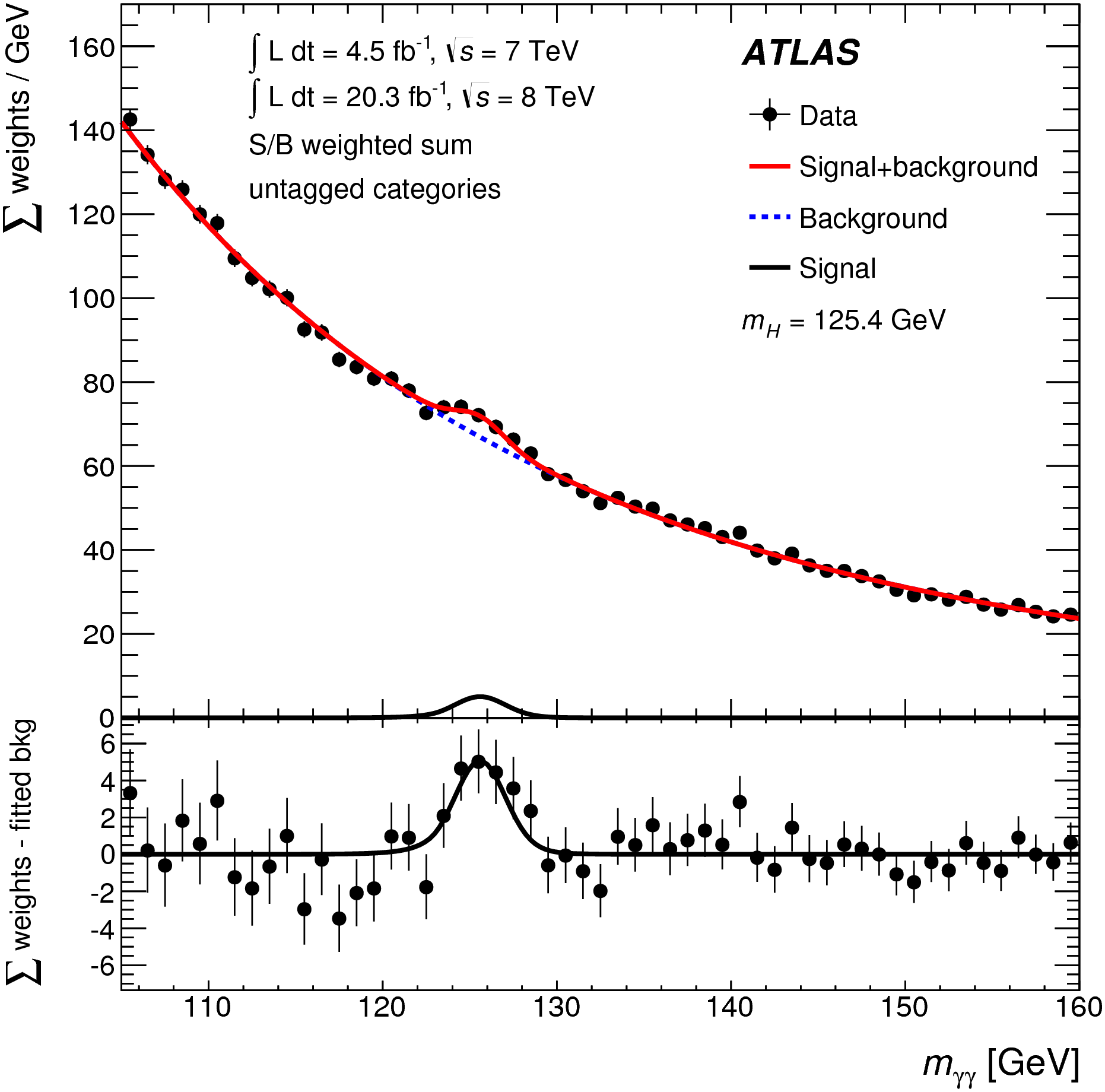} \hfill
\includegraphics[width=0.49\textwidth,height=6cm]{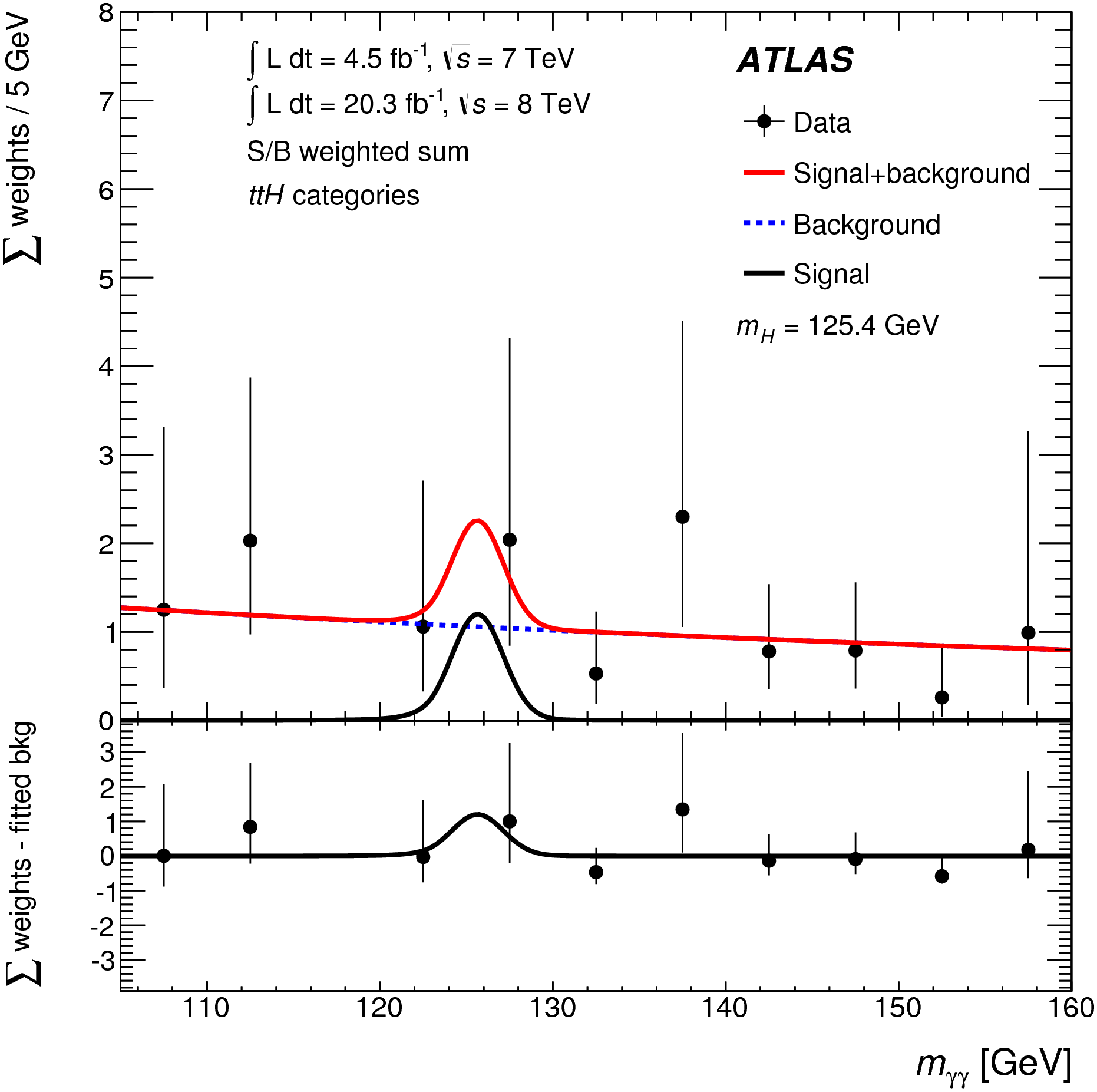} 
\vspace*{-0.5cm}
\caption{
Diphoton invariant mass spectra observed in the 7 TeV and 8 TeV data.
Left: all Higgs boson production modes (untagged). Right: ttH production mode.
In each plot the contribution from the different 
categories in each group is weighted according to the S/B ratio in each category. 
The error bars represent 68\% confidence intervals of the weighted sums. 
The solid red line indicates the fitted signal plus background model when the 
Higgs boson mass is fixed at $m_{\rm H} =125.4$~GeV. The background component of each 
fit is shown with a dotted blue line. Both the signal plus background and 
background-only curves reported here are obtained from the sum of the individual 
curves in each category weighted by their signal-to-background ratio. 
The bottom plot in each figure represents the data relative to the background component 
of the fitted model. 
}
\label{fig:gammagammalep1}
\vspace*{-0.5cm}
\end{figure}

\clearpage
Regarding the ATLAS Higgs diphoton searches, the Run-1 results show a clear peak
in the invariant mass spectrum for all Higgs boson production modes, while the 
ttH production mode sensitivity is limited by the small statistics, as shown
in Fig.~\ref{fig:gammagammalep1} (from~\cite{gammagammarun1}).

\section{The SM ttH Production and Higgs Boson Decays}
\begin{figure}[h!]
\vspace*{-0.4cm}
\includegraphics[width=0.32\textwidth,height=5cm]{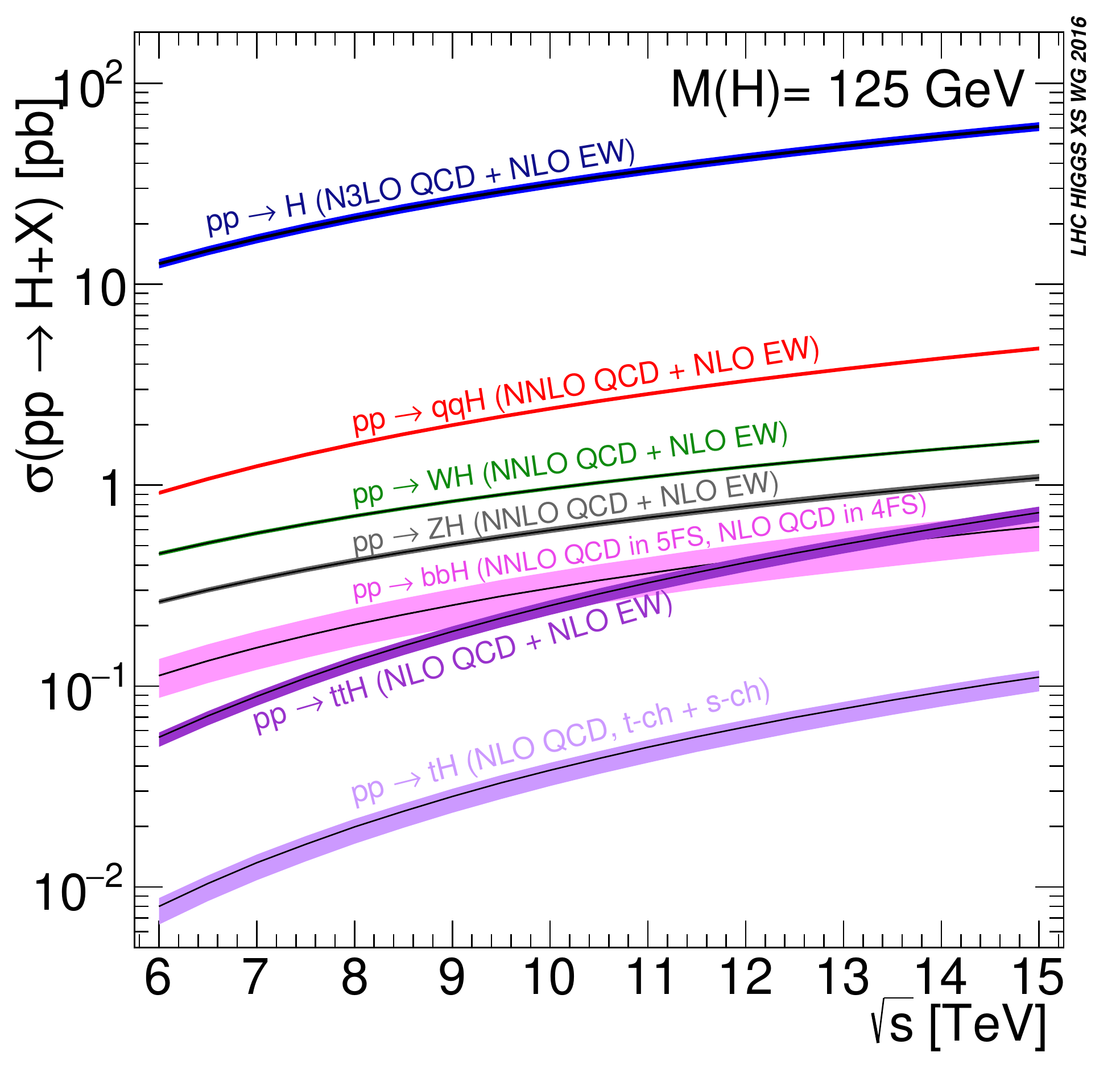}\hfill
\includegraphics[width=0.32\textwidth,height=5cm]{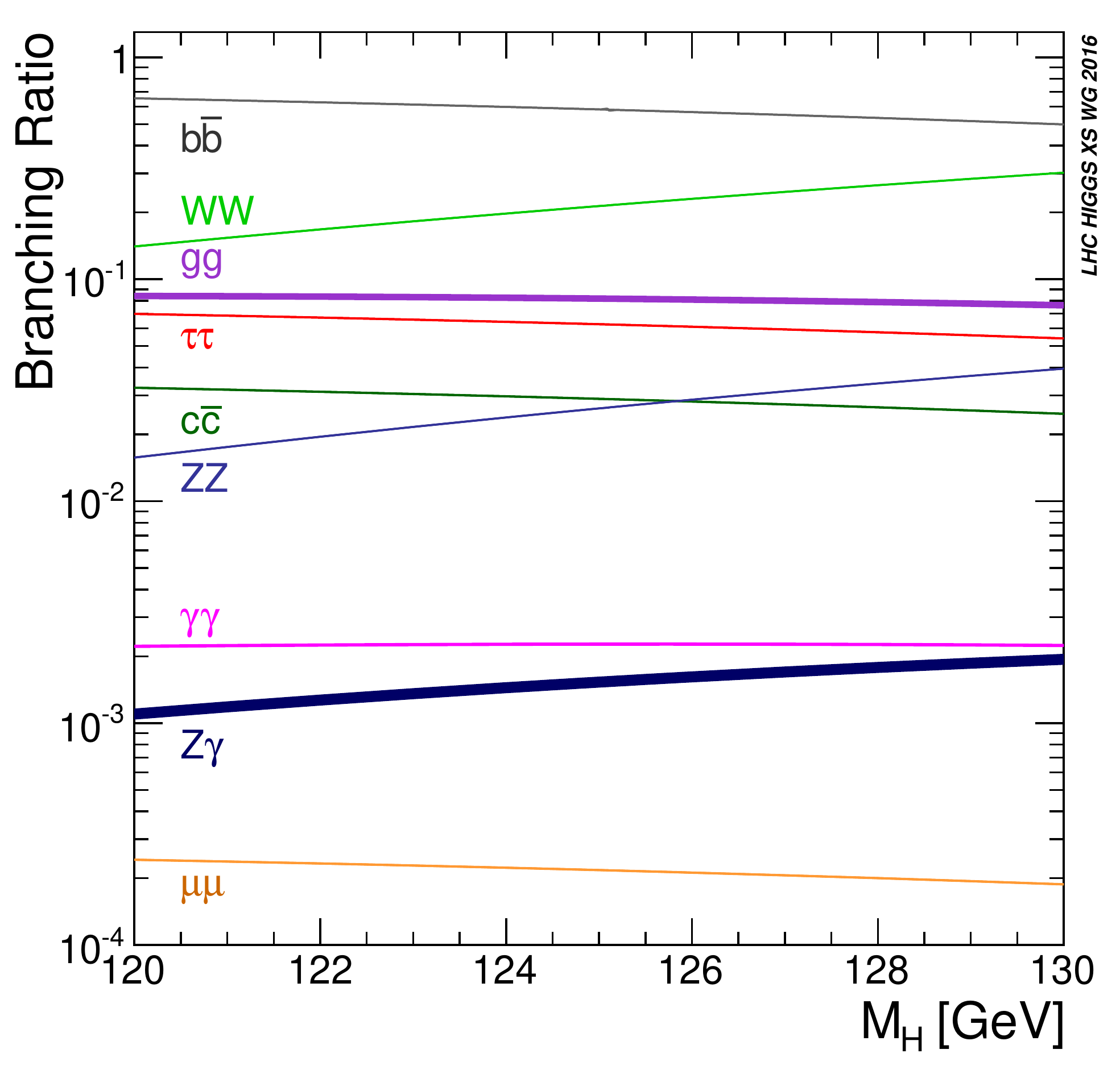}\hfill
\includegraphics[width=0.32\textwidth,height=4.5cm,trim=0 -10mm 0 0]{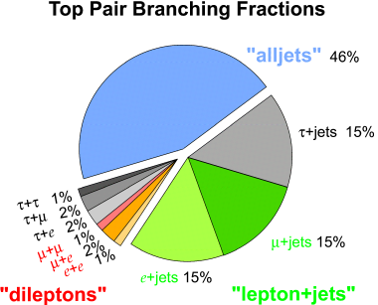}
\vspace*{-0.5cm}
\caption{
Left: SM Higgs boson production cross-sections as a function of 
      the LHC center-of-mass energy.
Center: SM Higgs boson branching ratios as a function of the Higgs boson mass.
Right: SM top-antitop branching ratios.
}
\label{fig:xsec}
\end{figure}

The ttH production cross-section, Higgs boson decay branching fractions  
and the top-antitop decay branching fractions are given in 
Fig.~\ref{fig:xsec} (from~\cite{deFlorian:2016spz}).
The ttH production cross-section increased by about a factor four from 
LHC Run-1 (8~TeV) to Run-2 (13~TeV).
A 125~GeV Higgs boson can decay into various pairs of particles.
This defines the signatures in the detector, together with the top-antitop decays
into alljets, lepton+jets and dileptons final states.

\begin{figure}[h!]
\vspace*{-0.6cm}
\includegraphics[width=0.5\textwidth,height=3cm,trim=0 -40mm 0 0]{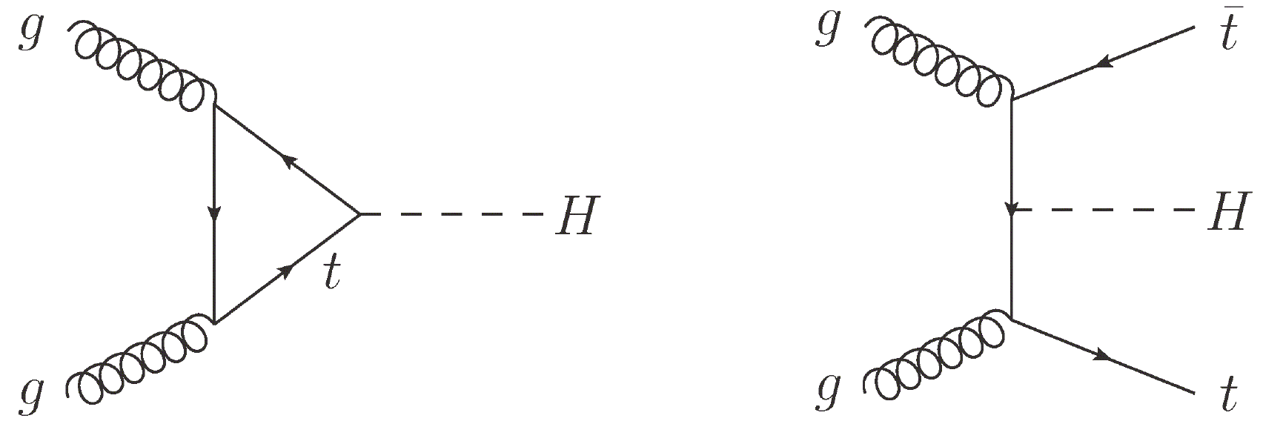} \hfill
\includegraphics[width=0.32\textwidth,height=5cm]{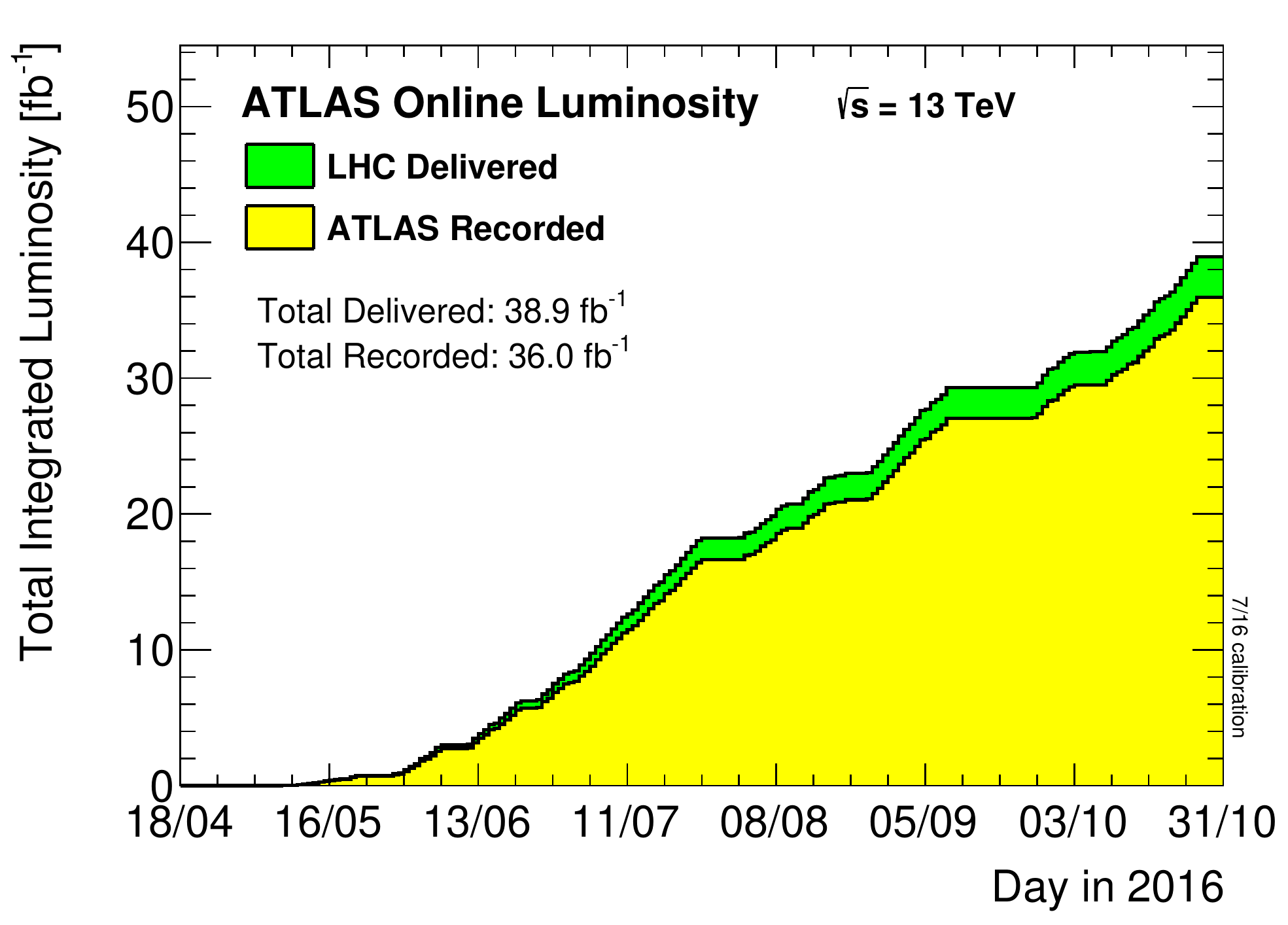}
\vspace*{-0.5cm}
\caption{
Feynman diagrams of Higgs boson production involving the ttH coupling.
Left: loop production.
Center: direct production.
Right: ATLAS 2016 delivered and recorded integrated luminosity.
}
\label{fig:feynman}
\vspace*{-0.5cm}
\end{figure}

Figure~\ref{fig:feynman} illustrates the Higgs boson production via a loop process 
involving the top Yukawa coupling, as well as the direct ttH production.
Although the highest Higgs boson production rate is via a loop ($\rm gg\rightarrow H$), 
the tree-level direct measurements of $\rm pp\rightarrow ttH$ are more sensitive to 
physics beyond the SM.

Some characteristics of the ttH analyses are given, ordered according 
to the Higgs boson decay modes:
\begin{itemize}
\item $\rm H\rightarrow WW, ZZ, \tau\tau$ produce multilepton events,
\item $\rm H\rightarrow \gamma\gamma$ 
      has a narrow bump in the diphoton invariant mass spectrum, and 
\item $\rm H\rightarrow bb$ has a large background rate.
\end{itemize}

\section{LHC Luminosity}

The measurements of the Higgs boson properties require high statistics datasets.
The LHC performance was excellent during the first phase of Run-2 in delivering proton-proton 
collisions, and ATLAS recorded them with high efficiency.
In 2015, 3.2~fb$^{-1}$, and in 2016, 36.0~fb$^{-1}$ data were recorded. The results presented 
here are based on the analysed data up to July 2016 corresponding to $3.2+10.0=13.2$~fb$^{-1}$.
Figure~\ref{fig:feynman} (from~\cite{LHC}) shows the increase of integrated luminosity 
delivered to and recorded by the ATLAS experiment.

For a luminosity of 13.2~fb$^{-1}$ and a ttH production cross-section of 
507.1~fb~\cite{deFlorian:2016spz}\footnote{computed at NLO in QCD and electroweak
couplings. It has uncertainties of $^{+5.8\%}_{-9.2\%}$ from QCD renormalization/factorization
scale choice and $\pm 3.6\%$ from parton distribution function uncertainties (including $\alpha_{\rm s}$
uncertainties).},
about 6700 ttH events were expected to be produced.
The experimental challenge was to detect these events in the about $10^{15}$ background 
events with a production cross-section of $78.1\pm 2.9$~mb~\cite{inelastic},
thus, finding one ttH event in about 154 billion background events.

\section{ttH (Multilepton) Agreement in Validation Regions, and Signal and Background Compositions}

\begin{figure}[h!]
\vspace*{-0.4cm}
\includegraphics[width=0.32\textwidth,height=5cm]{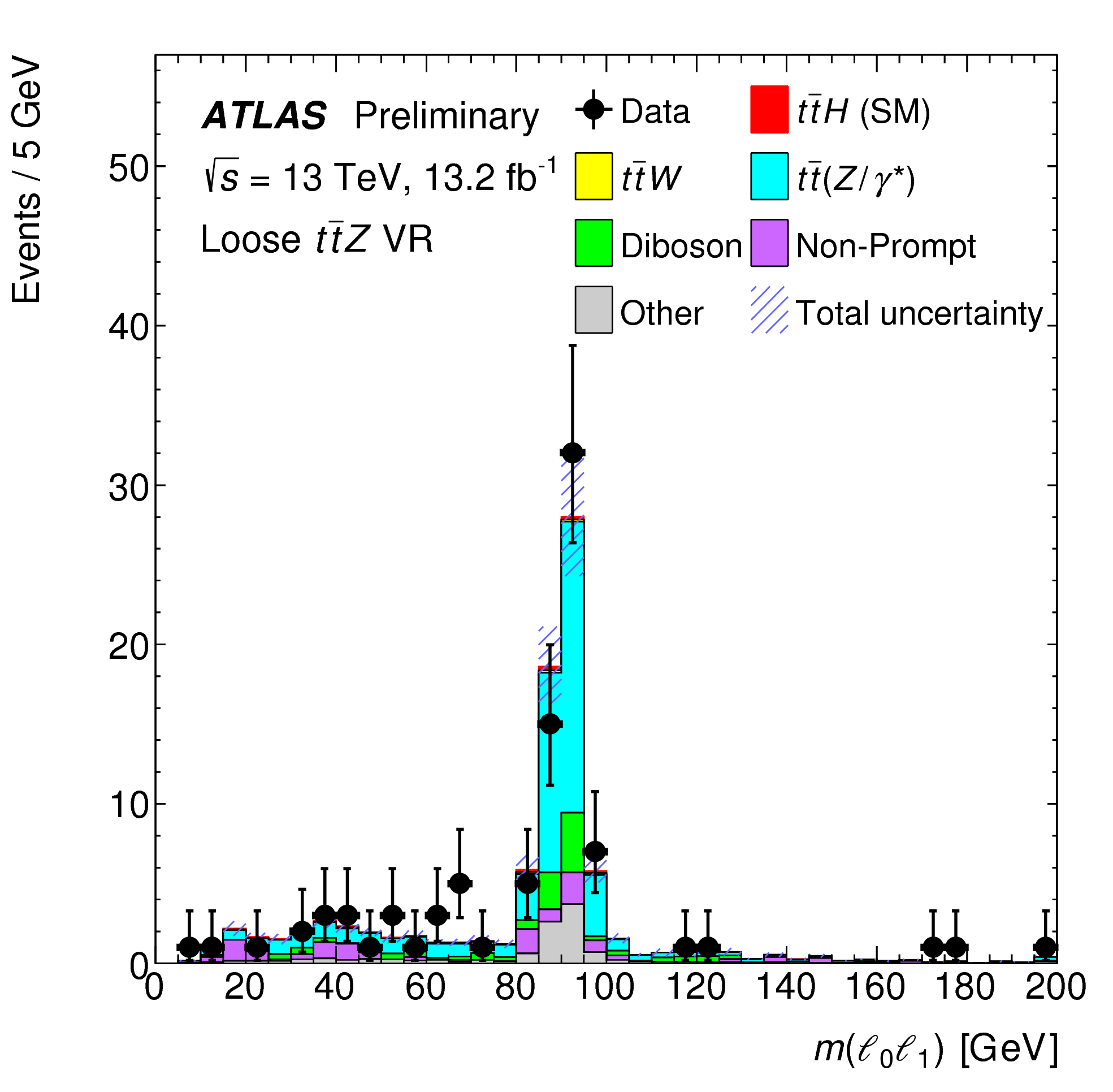}\hfill
\includegraphics[width=0.32\textwidth,height=5cm]{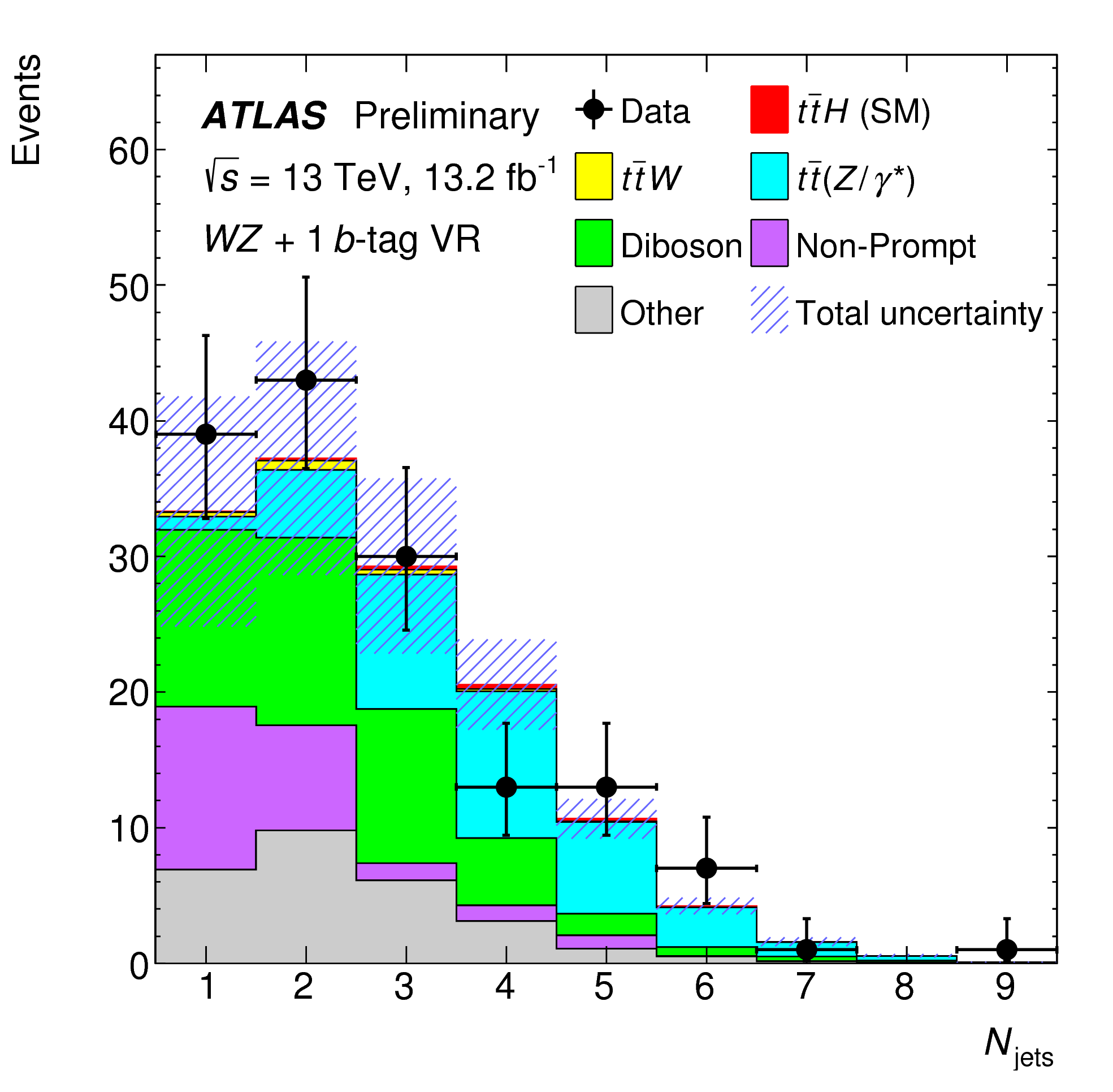}\hfill
\includegraphics[width=0.32\textwidth,height=5cm]{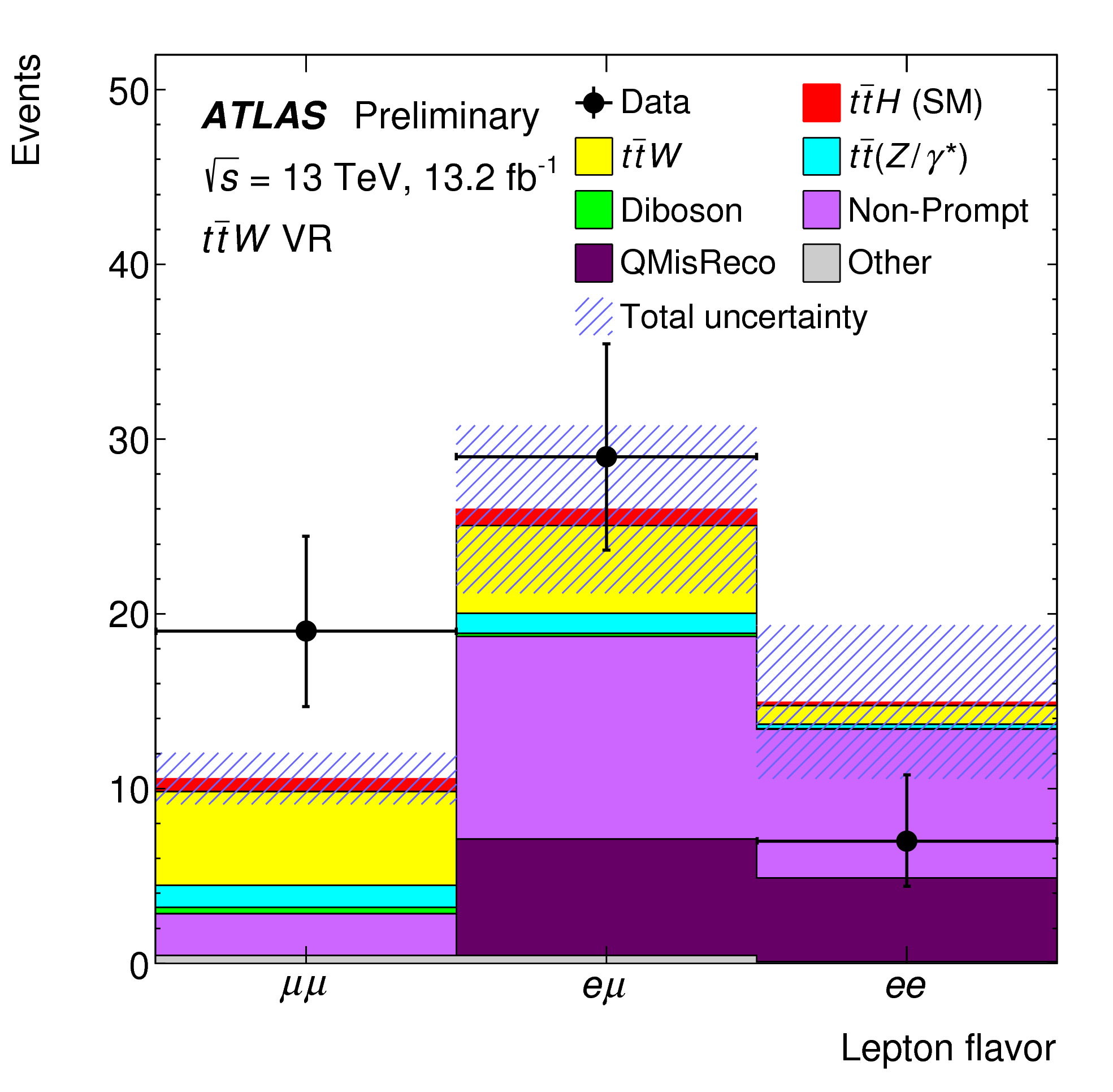}
\vspace*{-0.5cm}
\caption{
Left: invariant mass of leptons 0 and 1 for a loose ttZ validation region. 
      The leptons are labeled in the same way as for the $3\ell$ signal region. 
      The events away from the Z peak are those satisfying the Z selection with 
      leptons 0 and 2.
Center: jet multiplicity in the WZ+1bjet validation region. 
Right: lepton flavor composition for events in the ttW validation region. 
      Non-prompt lepton and charge mis-reconstruction backgrounds 
      (indicated as ``QMisReco'') are estimated using data as described in 
      Section 6.2 of Ref.~\cite{multilepton}.
}
\label{fig:validation}
\end{figure}

In order to search for a small number of signal events, it is essential to understand the 
background reactions with precision. The background reactions were studied in 
dedicated validation regions which are orthogonal to the signal regions. 
The validation regions are close to the signal regions and demonstrate 
the good agreement between the observed and simulated ttZ, WZ+1bjet, and ttW events, 
as shown in Fig.~\ref{fig:validation} (from~\cite{multilepton}), and 
listed in Table~\ref{tab:validation} (from~\cite{multilepton}).

\begin{figure}[h!]
\begin{center}
\includegraphics[width=0.58\textwidth,height=7.1cm]{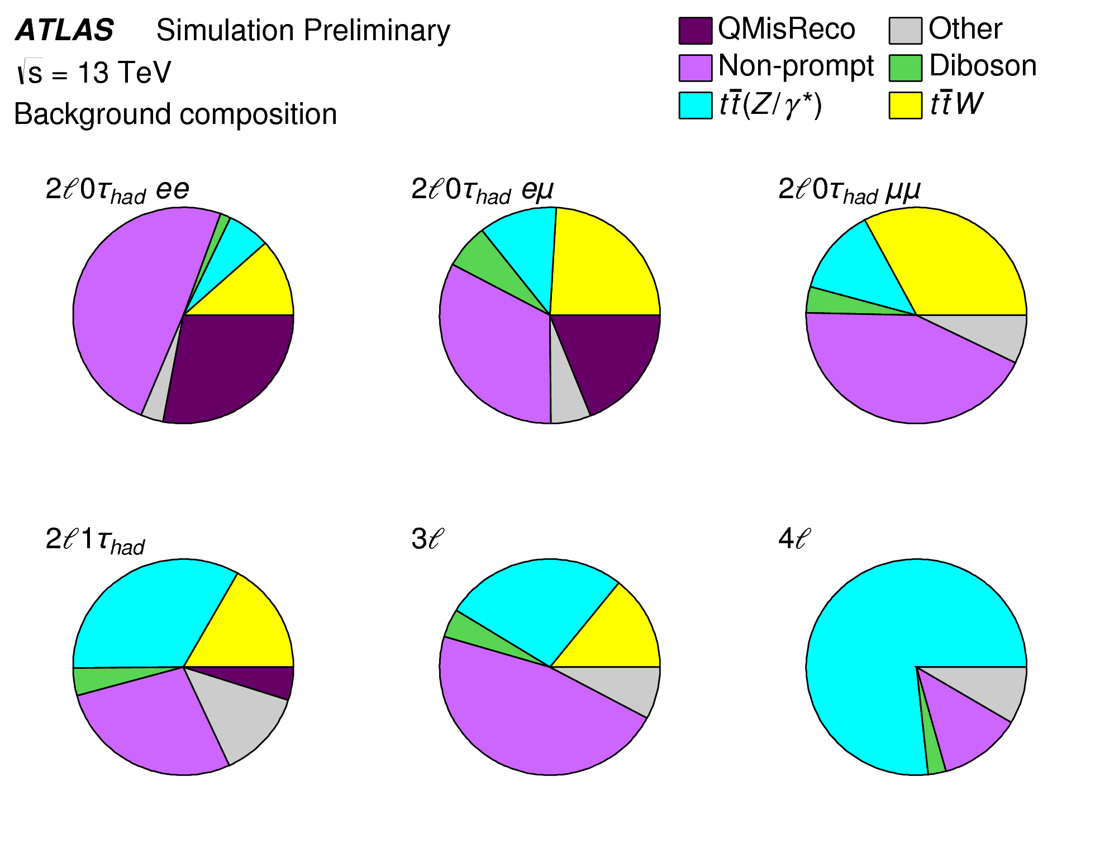} 
\end{center}
\vspace*{-1.2cm}
\caption{
Expected contribution to the background in each channel from various sources, using values of the background estimates before the fit. Charge mis-reconstruction backgrounds are indicated as ``QMisReco''.
}
\label{fig:composition}
\vspace*{-0.5cm}
\end{figure}

The signal compositions (Higgs boson decay modes) in the four multilepton channels
are given in Table~\ref{tab:composition} (from~\cite{multilepton}).
For example, the simulations
show that in the $2\ell1\tau_{\rm had}$ category, the $\rm H\rightarrow \tau\tau$ decay is
enriched with 51\% contribution.
The background compositions in six categories of the four cut-and-counting analyses 
are illustrated in Fig.~\ref{fig:composition} (from~\cite{multilepton}). 

\begin{table}[htb]
\vspace*{-0.2cm}
\caption{
Expected and observed event yields in validation regions (VR). 
The quoted uncertainties in the expectations include all systematic uncertainties. 
``Purity'' indicates the fraction of events in the VR expected to arise from the 
targeted process (ttZ for the first VR, WZ for the second, and ttW for the third).
\label{tab:validation} }
\begin{center} 
\vspace*{-0.1cm}
\begin{tabular}{cccc} \hline
VR        & Purity (\%) & Expected  & Data \\ \hline
Loose ttZ & 58          & $91\pm12$ & 89   \\
WZ+1bjet  & 33          & $137\pm27$& 147  \\
ttW       & 22          & $51\pm10$ & 55   \\ \hline
\end{tabular} 
\end{center}
\end{table}

\begin{table}[htb]
\vspace*{-0.4cm}
\caption{
Fraction of the expected ttH signal arising from different Higgs boson decay modes 
in each analysis category, and acceptance ($A$) times efficiency ($\epsilon$).
The decays contributing to the ``other'' column are 
dominantly $\rm H\rightarrow \mu\mu$ and $\rm H\rightarrow bb$. 
Rows may not add to 100\% due to rounding. 
The acceptance times efficiency includes Higgs boson and top quark branching fractions, 
detector acceptance, and reconstruction and selection efficiency, 
being computed relative to inclusive ttH production.
\label{tab:composition} }
\vspace*{-0.2cm}
{
\begin{center}
\begin{tabular}{cccccc} \hline
          &\multicolumn{4}{c}{Higgs\,boson\,decay\,mode\,(\%)}    & $A\times\epsilon$ \\
Category               & WW$^*$  & $\tau\tau$  & ZZ$^*$ & Other & $(\times 10^{-4})$\\ \hline
$2\ell0\tau_{\rm had}$ & 77      & 17          & 3      & 3     & 14   \\ 
$2\ell1\tau_{\rm had}$ & 46      & 51          & 2      & 1     & 2.2  \\
$3\ell$                & 74      & 20          & 4      & 2     & 9.2  \\ 
$4\ell$                & 72      & 18          & 9      & 2     & 0.88 \\\hline
\end{tabular}

\end{center}
}
\vspace*{-0.3cm}
\end{table}

\clearpage
\section{Two Same-charge Light Leptons Plus One Hadronic Tau Final State}

\begin{figure}[h!]
\vspace*{-0.4cm}
\includegraphics[width=0.49\textwidth,height=5cm]{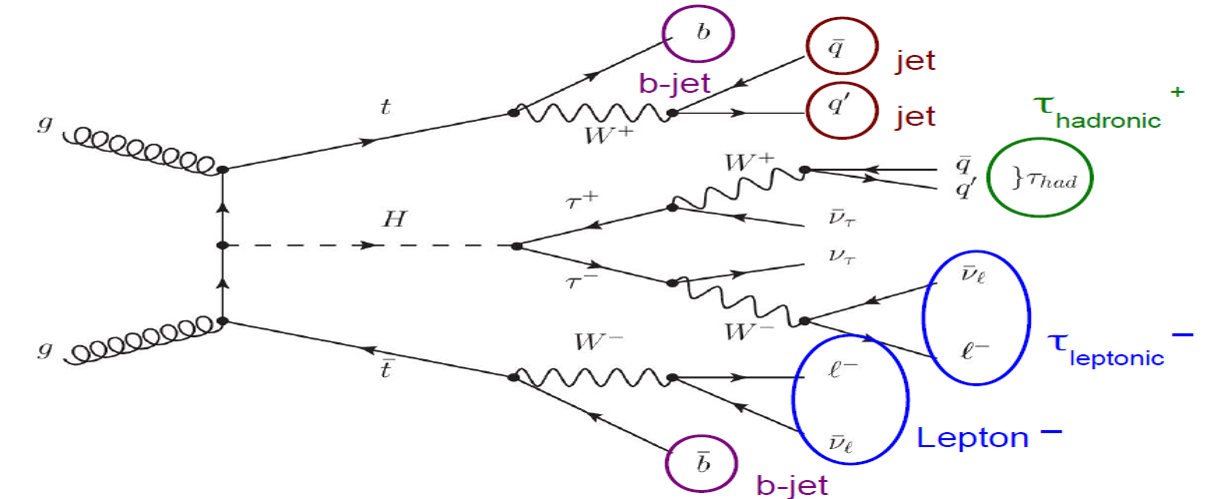} \hfill
\includegraphics[width=0.49\textwidth,height=5cm]{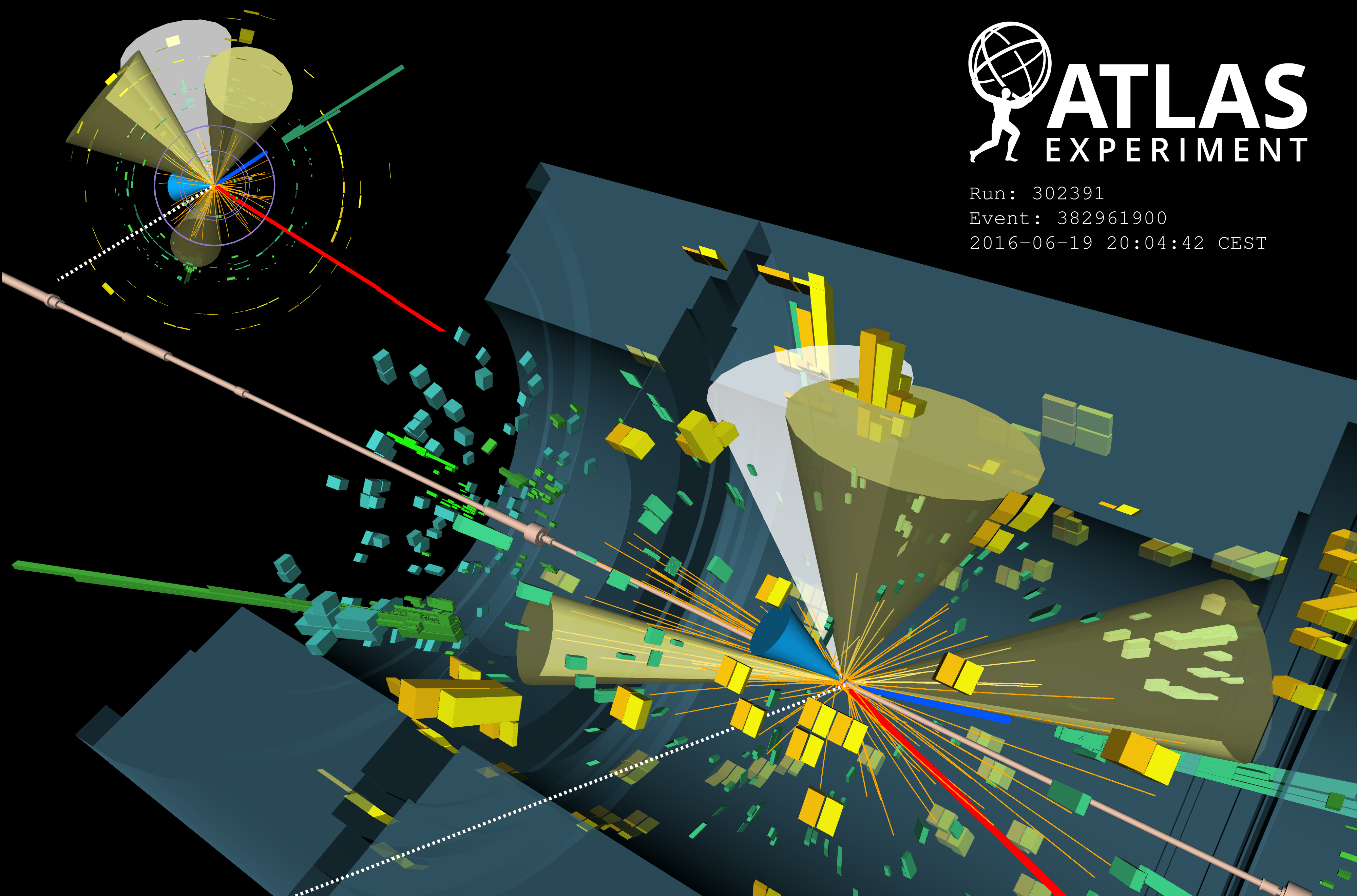} 
\vspace*{-0.5cm}
\caption{
Left: Feynman diagram of the two same-charge leptons plus one hadronic tau analysis channel
($2\ell1\tau_{\rm had}$ category).
Right: event display for a candidate e$\mu\tau_{\rm had}$ event in the 
$2\ell1\tau_{\rm had}$ category.
The blue track is the selected electron; the red track is the selected muon; 
and the white cone is the $\tau_{\rm had}$ candidate. The azure cone is the selected b-tagged jet, 
and the three yellow cones are the non-b-tagged jets. Green and yellow bars 
indicate energy deposits in the electromagnetic (liquid argon) and hadronic (tile) 
calorimeters, respectively. 
}
\label{fig:feynmanchannel}
\vspace*{-0.5cm}
\end{figure}

In the two same-charge light leptons plus one hadronic tau analysis channel, as illustrated 
in Fig.~\ref{fig:feynmanchannel}, one light lepton arises 
from the tau decay, and one from the top decay. 
The other tau and the other top decay hadronically, leading to the signature 
with multijets and b-jets. A selected candidate event display is also shown in
Fig.~\ref{fig:feynmanchannel} (from~\cite{multilepton}).

\section{Two Same-charge Light Leptons Without Hadronic Tau Final State, Three and Four Light Leptons Final States}

\begin{figure}[h!]
\vspace*{-0.4cm}
\includegraphics[width=0.49\textwidth,height=5cm]{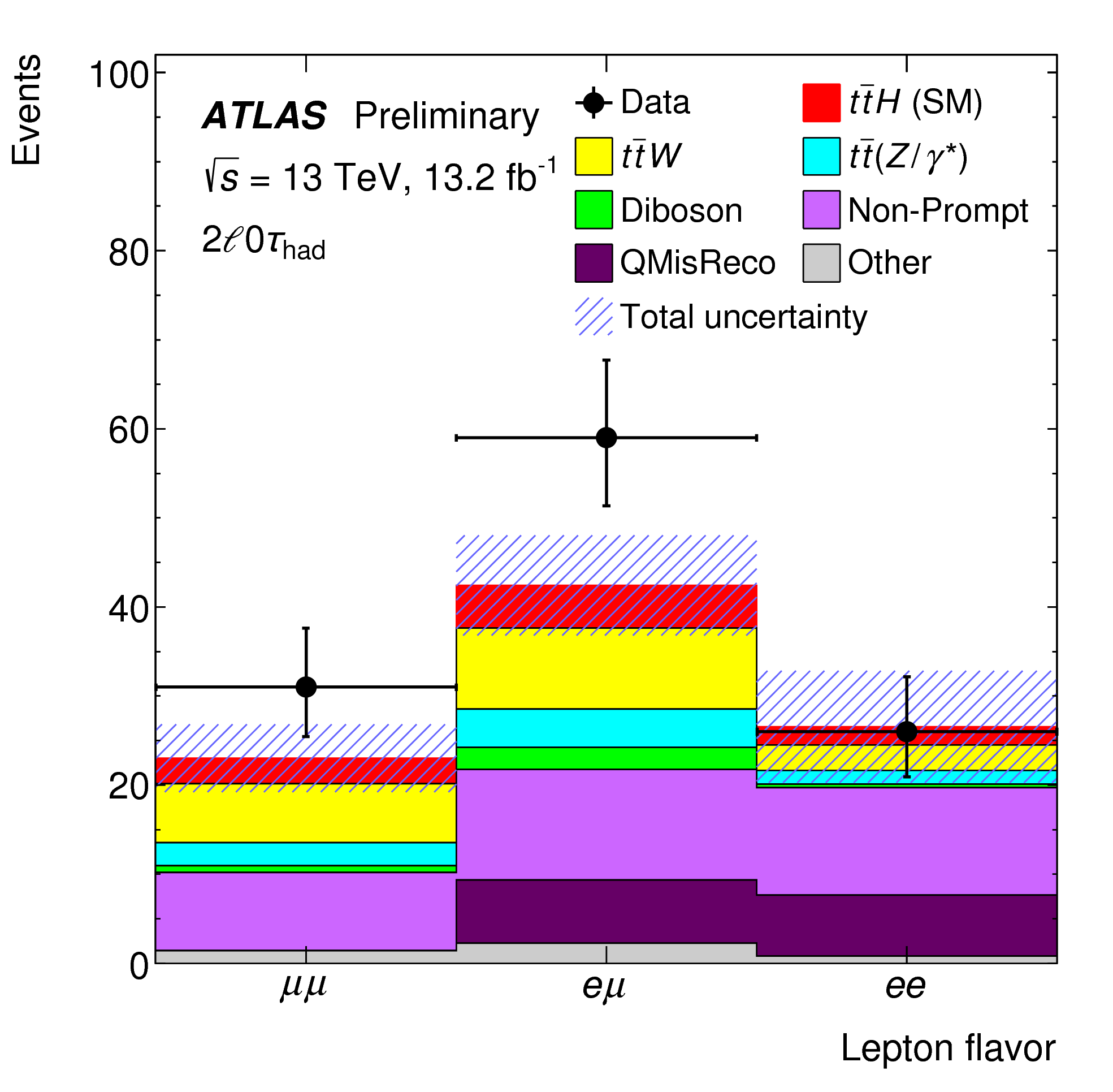} \hfill
\includegraphics[width=0.49\textwidth,height=5cm]{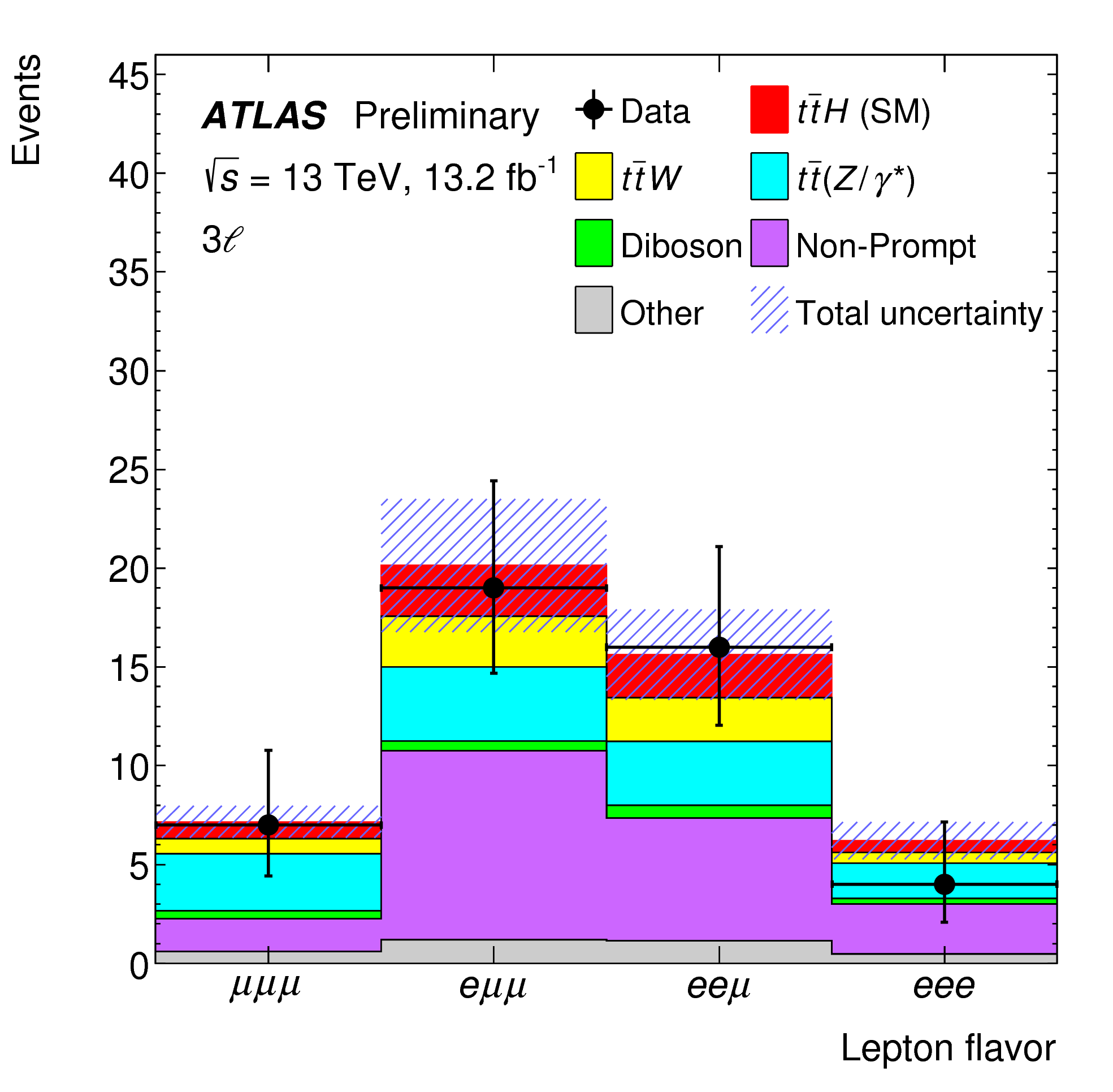} 
\vspace*{-0.5cm}
\caption{
Left: 
lepton flavor composition in the $2\ell0\tau_{\rm had}$ signal region.
The signal is set to the SM expectation ($\mu_{\rm ttH} = 1$) 
while the background expectation is pre-fit (using initial values of the background systematic 
uncertainty nuisance parameters). The hatched region shows the total uncertainty 
on the background plus SM signal prediction in each bin. 
Charge mis-reconstruction backgrounds are indicated as ``QMisReco''. 
Right:
lepton flavor composition in the $3\ell$ signal region. 
The signal is set to the SM expectation ($\mu_{\rm ttH} = 1$) and the background expectation 
is pre-fit (using initial values of the background systematic 
uncertainty nuisance parameters). The hatched region shows the total uncertainty 
on the background plus SM signal prediction in each bin.
}
\label{fig:2and3leptons}
\vspace*{-0.5cm}
\end{figure}

For the ttH search with two same-charge light leptons without hadronic tau final state, 
and three leptons, Fig.~\ref{fig:2and3leptons} (from~\cite{multilepton}) 
shows the number of data, SM signal and background composition.
For the four light leptons final state, no data event passes the selection.

\section{ttH (Multilepton) Results}

\begin{figure}[h!]
\vspace*{-0.4cm}
\includegraphics[width=0.49\textwidth,height=5cm]{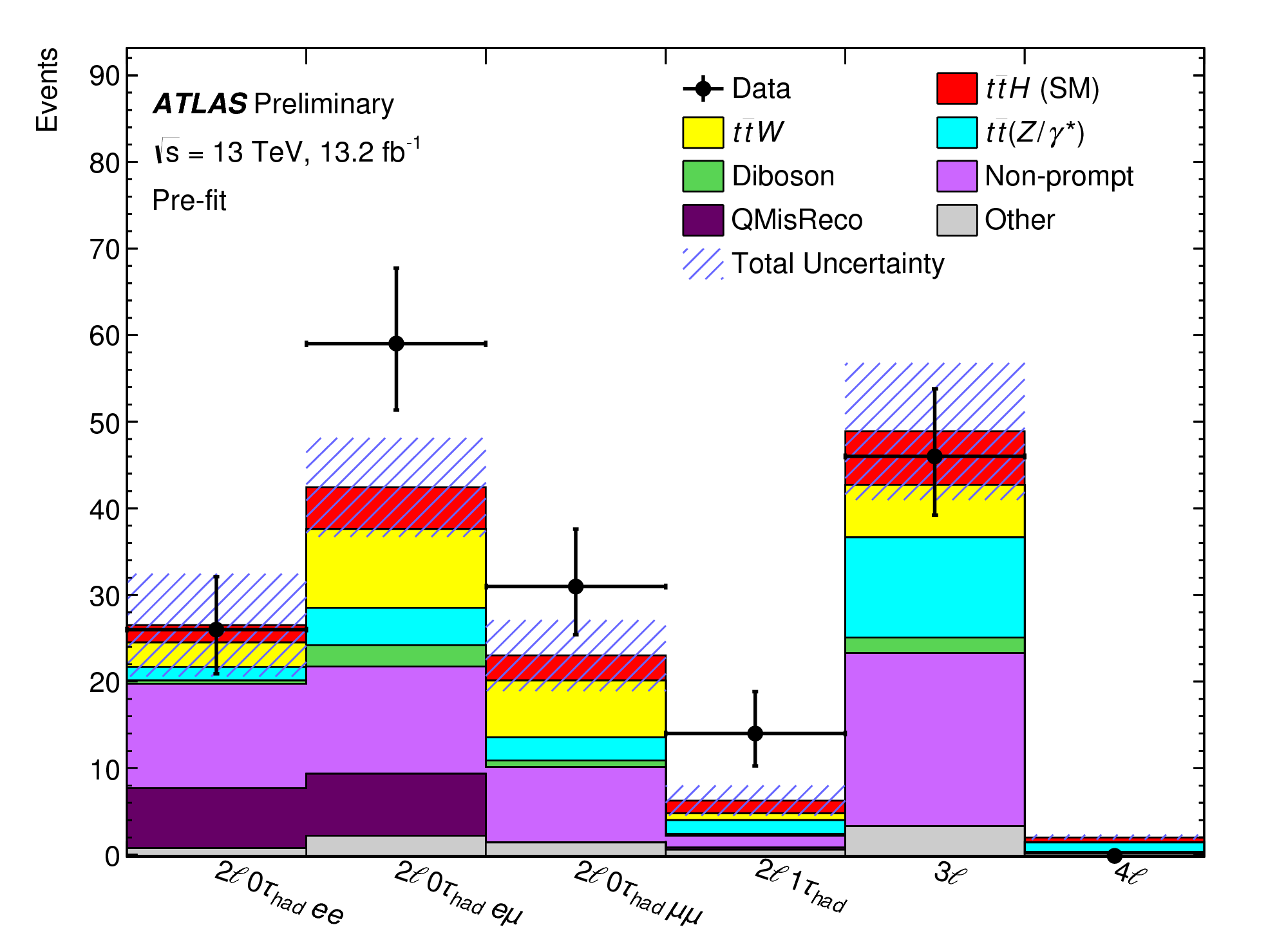} \hfill
\includegraphics[width=0.49\textwidth,height=5cm]{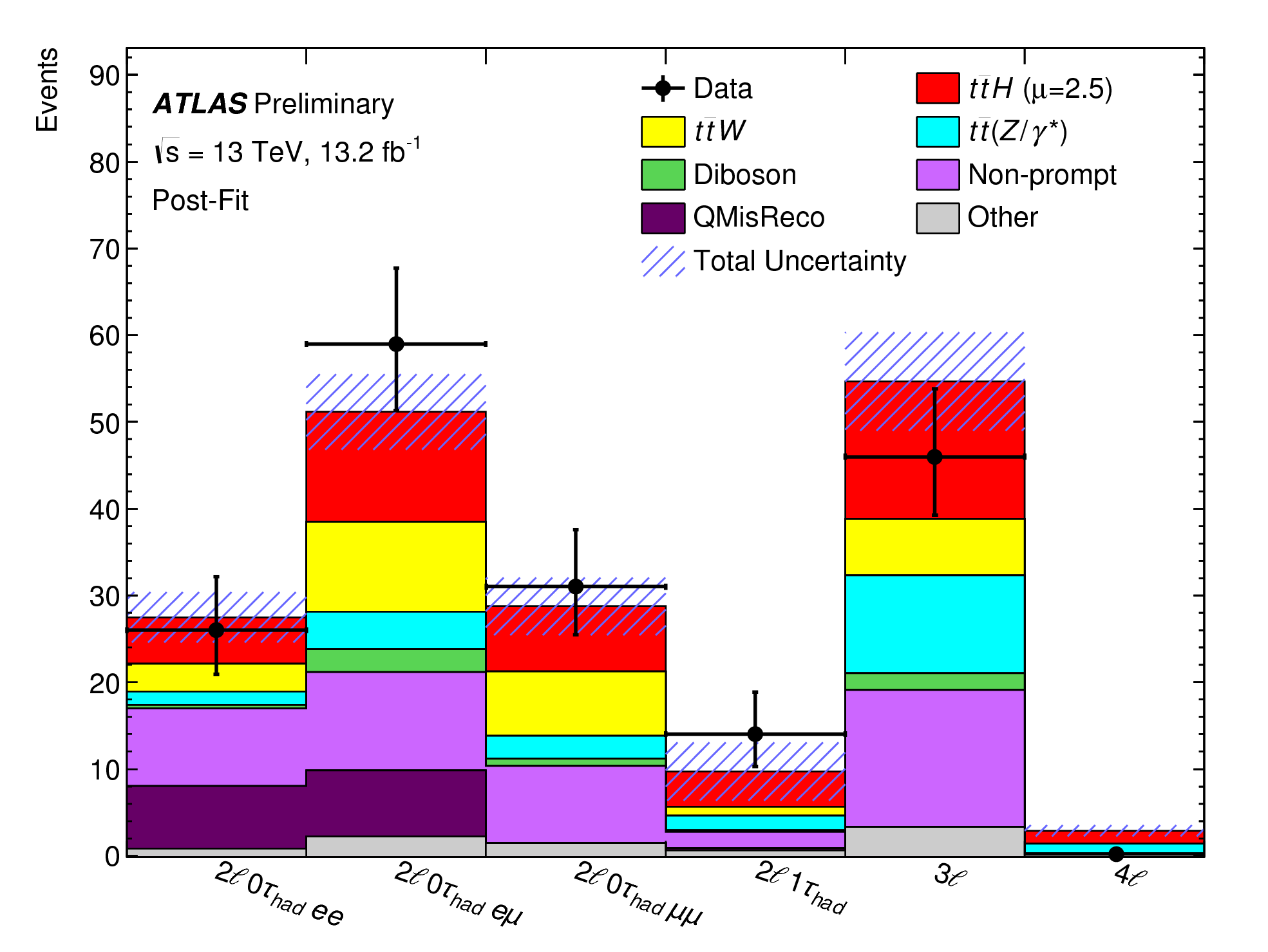} 
\vspace*{-0.5cm}
\caption{
Left: 
pre-fit background and signal predictions along with observed data yields for each signal region. 
The ttH prediction corresponds to the SM expectation ($\mu_{\rm ttH} = 1$). 
The charge mis-reconstruction backgrounds are indicated as ``QMisReco''. 
Right:
post-fit background and signal predictions as well as observed data yields for each signal region. 
The background expectations have been updated to reflect the values of systematic uncertainty 
nuisance parameters after the fit to data. The ttH prediction corresponds to the best-fit 
value ($\mu _{\rm ttH} = 2.5^{+1.3}_{-1.1}$) and the displayed total uncertainties reflect 
the uncertainty in ttH as well as the backgrounds. 
}
\label{fig:multileptonresults}
\vspace*{-0.5cm}
\end{figure}

\begin{figure}[h!]
\vspace*{-0.4cm}
\includegraphics[width=0.49\textwidth,height=5cm]{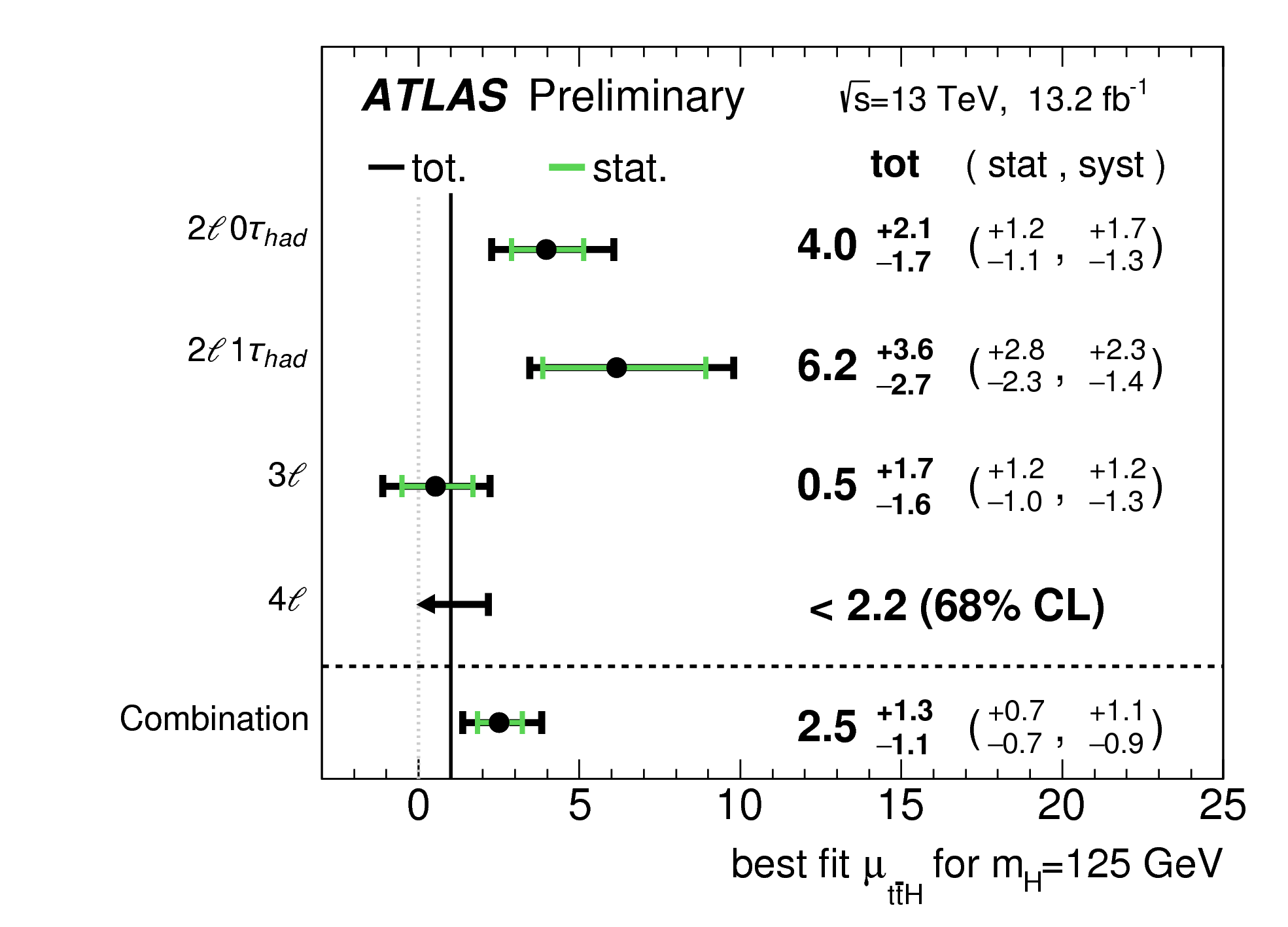} \hfill
\includegraphics[width=0.49\textwidth,height=5cm]{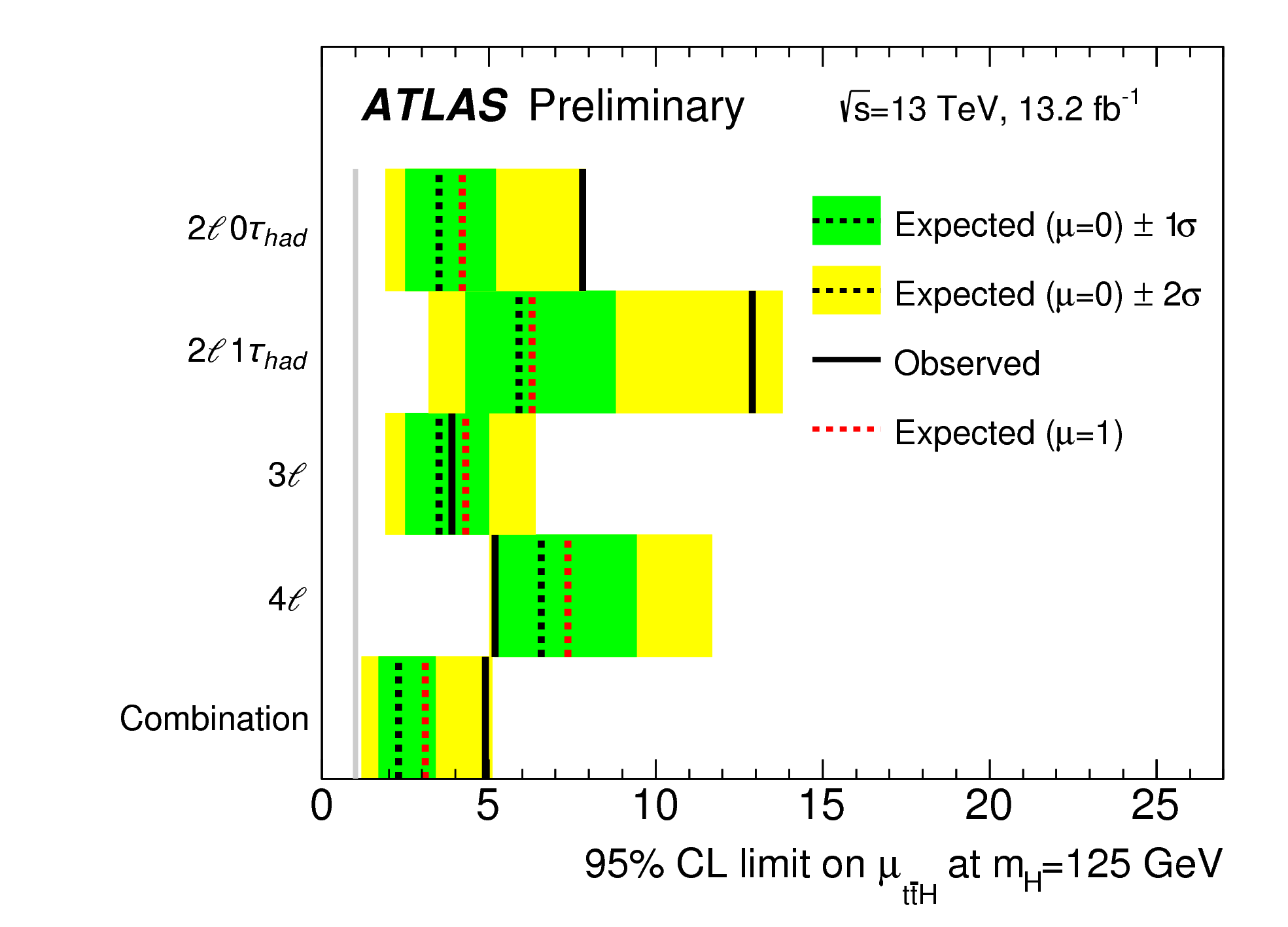} 
\vspace*{-0.5cm}
\caption{
Left: 
best fit values of the ttH signal strength $\mu_{\rm ttH}$ by final state category 
and combined. The SM prediction is $\mu_{\rm ttH}=1$.
For the $4\ell$ category, as zero 
events are observed, a 68\% CLs upper limit is shown instead. 
Right:
upper limits on the ttH signal strength μttH at 95\% CL by final state 
category and combined. The SM prediction is $\mu_{\rm ttH}=1$. 
The median upper limit that would be set in the presence of a 
SM ttH signal ($\mu=1$) is also presented. 
}
\label{fig:multileptonlimits}
\vspace*{-0.7cm}
\end{figure}

The results of the ttH search can be expressed as numbers of selected data, simulated signal 
and determined background events, as shown in Fig.~\ref{fig:multileptonresults} 
(from~\cite{multilepton}) for pre-fit and post-fit, and pre-fit
details are presented in Table~\ref{tab:tth} (from~\cite{multilepton}).
The post-fit gives $\mu_{\rm ttH} = 2.5^{+1.3}_{-1.1}$ indicating a mild data excess
compared to the SM prediction. 
The comparison of the obtained signal strengths $\mu_{\rm ttH}$ and 
their upper limits are given in Fig.~\ref{fig:multileptonlimits} (from~\cite{multilepton}).

\newcommand{\ttH}{\ensuremath{ttH}}
\newcommand{\tauh}{\ensuremath{\tau_\mathrm{had}}}
\newcommand{\twol}{\ensuremath{2\ell 0\tauh}}
\newcommand{\twoltau}{\ensuremath{2\ell 1\tauh}}
\newcommand{\threel}{\ensuremath{3\ell}}
\newcommand{\fourl}{\ensuremath{4\ell}}
\newcommand{\ttW}{\ensuremath{ttW}}
\newcommand{\phz}{\phantom{0}}

\begin{table}[h!]
\vspace*{-0.5cm}
\begin{center} 
 \caption{\label{tab:tth}
          Expected and observed yields in the six signal region 
          categories in 13.2~fb$^{-1}$ of data at $\sqrt{s} =$ 13~TeV. 
          The uncertainties in the background expectations due to systematic effects and 
          MC statistics are also shown. ``Other'' backgrounds include 
          $\rm tZ$, $\rm tWZ$, $\rm tHqb$, $\rm tHW$, $\rm tttt$, $\rm ttWW$, 
          and triboson production. Values are obtained pre-fit, i.e., 
          using the initial values of background systematic uncertainty 
          nuisance parameters.} 
\vspace{1mm}
{\small 
\renewcommand{\arraystretch}{0.8} 
 \begin{tabular}{lcccccc} 
  \hline
                     & \twol $\rm ee$      & \twol $\rm e\mu$  & \twol $\mu\mu$ & \twoltau        & \threel       & \fourl \\
                      \hline 
ttW                  & 2.9 $\pm$ 0.7   & 9.1 $\pm$ 2.5 & 6.6 $\pm$ 1.6  & 0.8 $\pm$ 0.4   & 6.1 $\pm$ 1.3 & --- \\ 
$\rm tt(Z/\gamma^*)$ & 1.55 $\pm$ 0.29 & 4.3 $\pm$ 0.9 & 2.6 $\pm$ 0.6  & 1.6 $\pm$ 0.4   & 11.5 $\pm$ 2.0\phz& 1.12 $\pm$ 0.20\\ 
Diboson               & 0.38 $\pm$ 0.25 & 2.5 $\pm$ 1.4 & 0.8 $\pm$ 0.5  & 0.20 $\pm$ 0.15 & 1.8 $\pm$ 1.0 & 0.04 $\pm$ 0.04\\ 
Non-prompt lepton\hspace*{-3mm}    & 12 $\pm$ 6\phz  & 12 $\pm$ 5\phz& 8.7 $\pm$ 3.4  & 1.3  $\pm$ 1.2  & 20 $\pm$ 6\phz& 0.18 $\pm$ 0.10\\ 
Ch.misID& 6.9 $\pm$ 1.3& 7.1 $\pm$ 1.7 & ---            & 0.24 $\pm$ 0.03 & ---          & --- \\ 
Other                 & 0.81 $\pm$ 0.22 & 2.2 $\pm$ 0.6 & 1.4 $\pm$ 0.4  & 0.63 $\pm$ 0.15 & 3.3 $\pm$ 0.8 & 0.12 $\pm$ 0.05 \\ 
\hline 
Total background      & 25 $\pm$ 6\phz  & 38 $\pm$ 6\phz& 20 $\pm$ 4\phz & 4.8 $\pm$ 1.4   & 43 $\pm$ 7\phz& 1.46 $\pm$ 0.25 \\ 
ttH (SM)             & 2.0 $\pm$ 0.5   & 4.8 $\pm$ 1.0 & 2.9 $\pm$ 0.6  & 1.43 $\pm$ 0.31 & 6.2 $\pm$ 1.1 & 0.59 $\pm$ 0.10 \\ 
\hline 
Data                  & 26              & 59            & 31             & 14               & 46           & 0\\ 
\hline
 \end{tabular} 
} 
\end{center} 
\vspace*{-0.6cm}
\end{table}

\clearpage
\section{ttH (\boldmath$\rm H\rightarrow \gamma\gamma$) Analysis}

\begin{figure}[h!]
\vspace*{-0.4cm}
\includegraphics[width=0.49\textwidth,height=6cm]{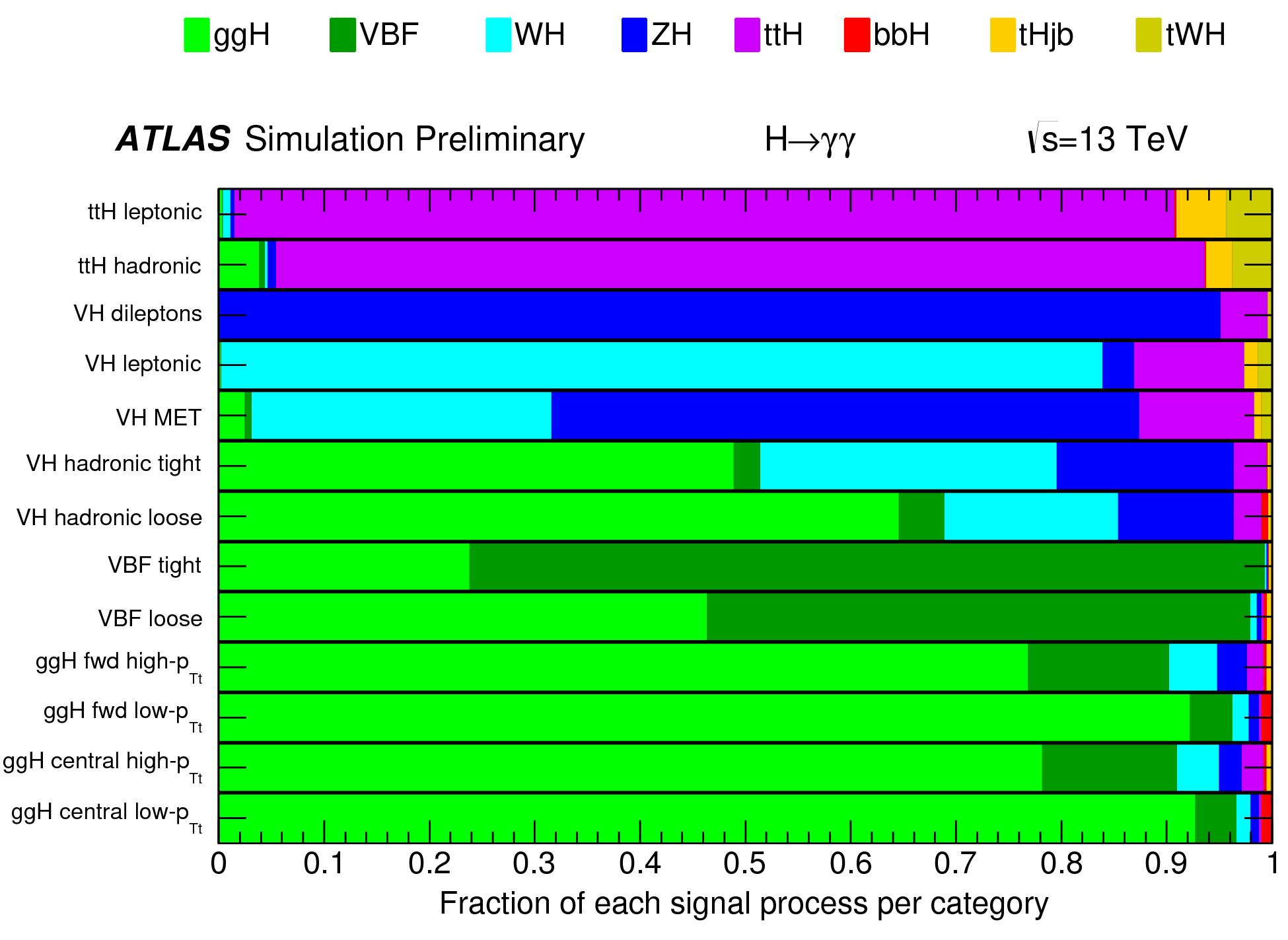} \hfill
\includegraphics[width=0.49\textwidth,height=6cm]{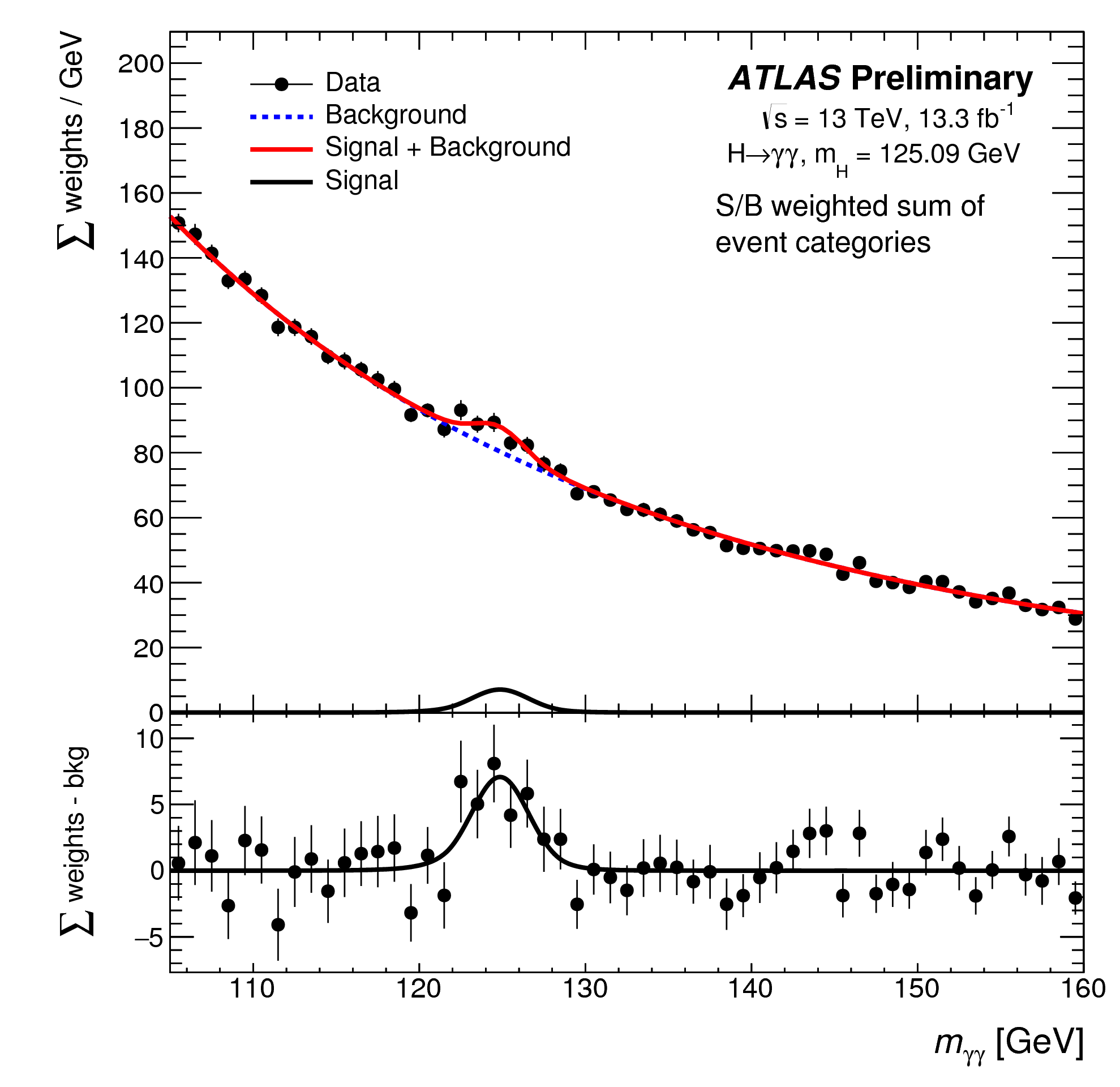} 
\vspace*{-0.5cm}
\caption{
Left: expected signal composition of the event categories for the full kinematic phase space.
Right:
invariant mass spectrum combining all production mode categories. 
The black data points indicate the measured distribution, where each event is 
weighted by the signal-to-background ratio of the event category it belongs to. 
The blue dashed curve represents the result of a background-only fit to the data, 
the red curve shows the signal+background distribution based on the fitted signal yields, 
while the black curve shows the signal component. 
The bottom inset displays the residuals of the data with respect to the fitted background 
component (bkg).
}
\label{fig:gg2modes}
\vspace*{-0.4cm}
\end{figure}

Owing to the excellent photon energy resolution of the ATLAS calorimeter,
the Higgs boson signal manifests itself as a narrow peak in the diphoton invariant mass 
spectrum on top of a smoothly falling background, and the Higgs boson
signal yield can be measured using an appropriate fit.
The search for ttH ($\rm H\rightarrow\gamma\gamma$)
in the initial LHC Run-2 data~\cite{gammagamma} follows the Run-1 strategy, and is performed in
two categories, 
one with associated leptons and 
one with associated hadrons. 
In both categories the ttH contents is above 90\% among all Higgs boson production modes,
as shown in Fig.~\ref{fig:gg2modes} (from~\cite{gammagamma}).

\begin{figure}[h!]
\vspace*{-0.4cm}
\includegraphics[width=0.49\textwidth,height=5cm]{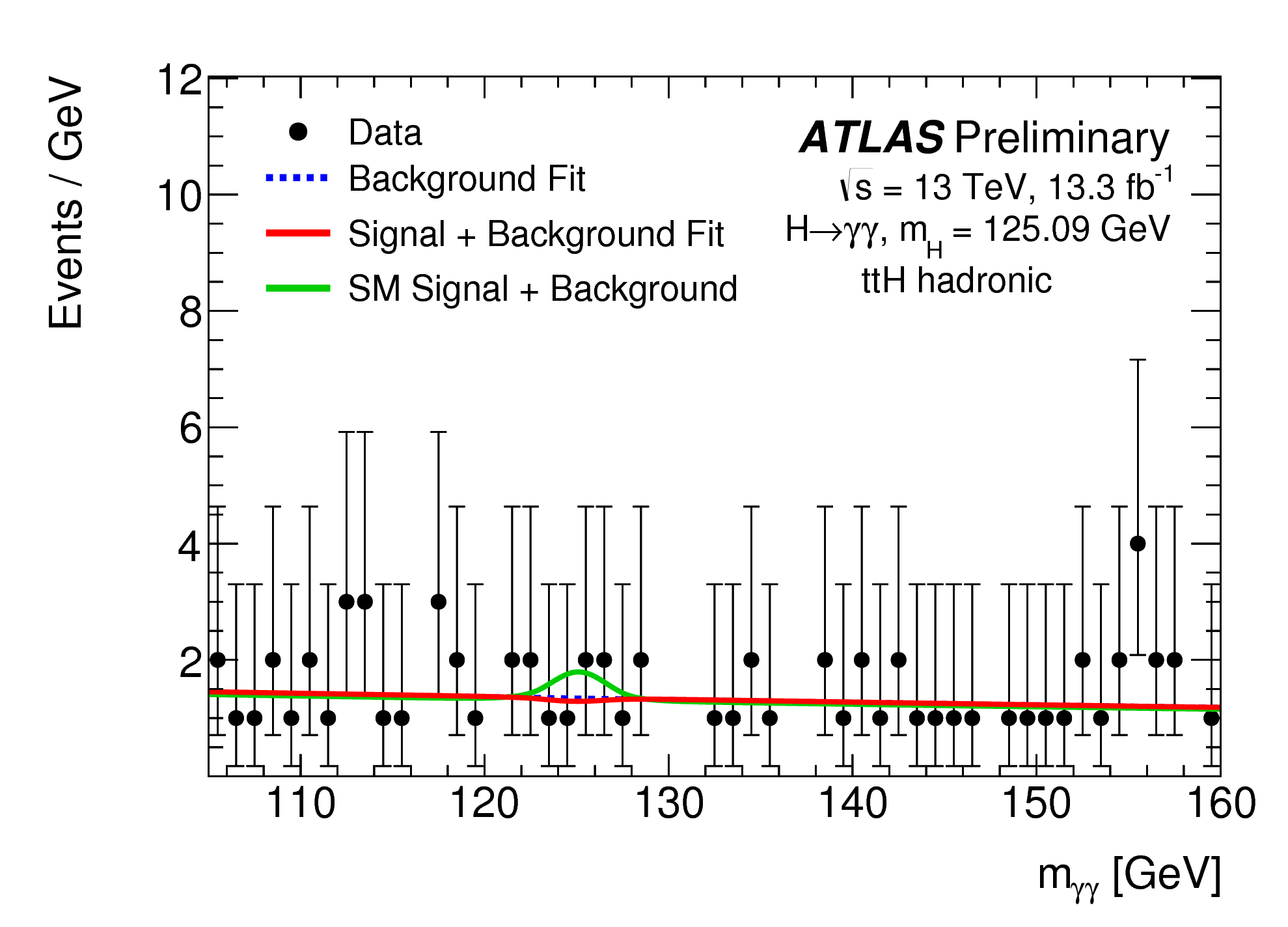} \hfill
\includegraphics[width=0.49\textwidth,height=5cm]{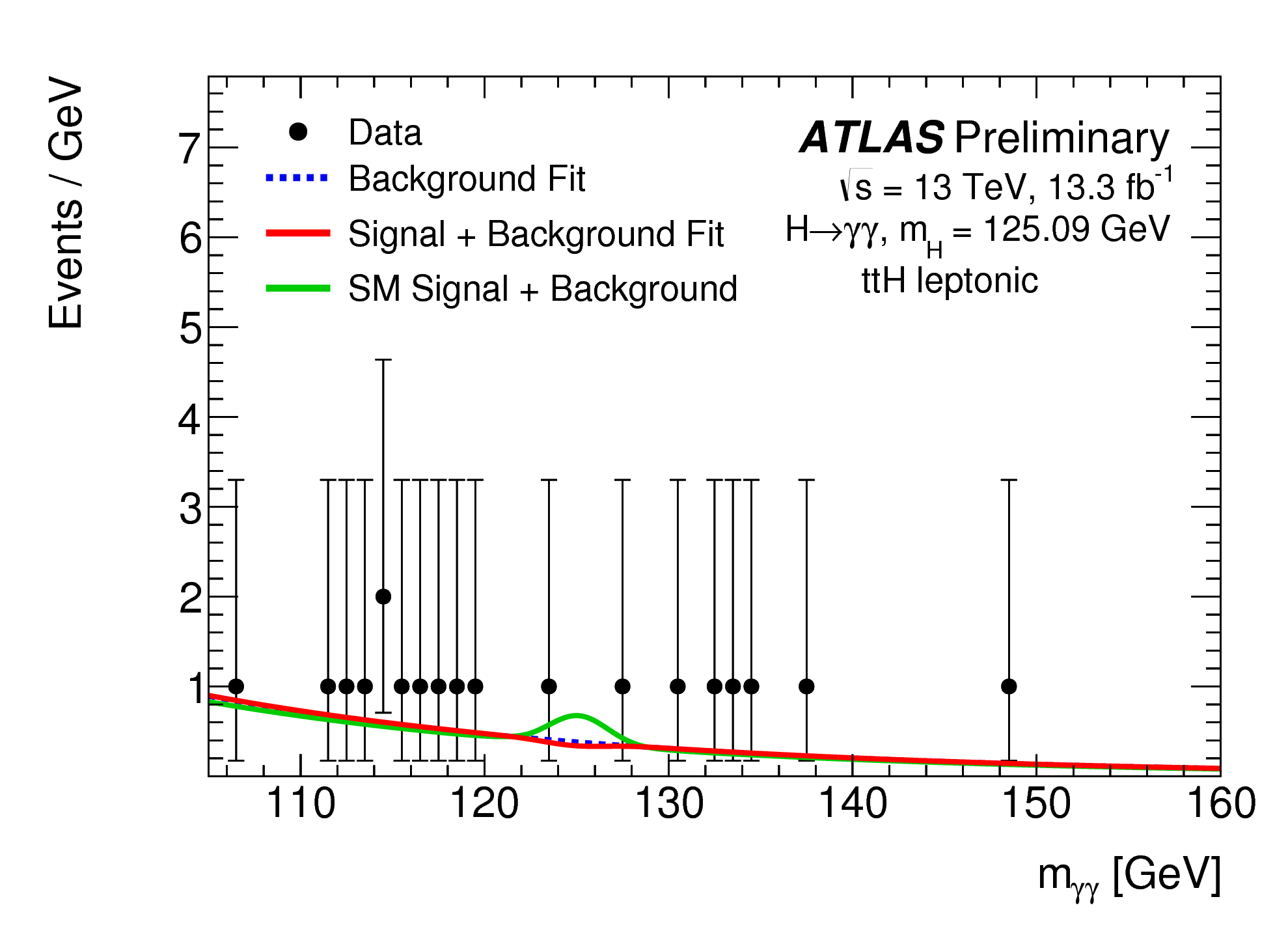}
\vspace*{-0.5cm}
\caption{
Invariant mass spectrum in the ttH ($\rm H\rightarrow \gamma\gamma$) search.
Left: hadronic production mode. 
Right: leptonic production mode. 
The black data points indicate the measured distribution, 
the blue dashed curve represents the result of a background-only fit to the data, 
the green curve marks the signal+background distribution based on the predicted 
SM signal for a Higgs boson mass of $m_{\rm H}= 125.09$~GeV, 
and the red curve shows the signal+background distribution based 
on the fitted signal yields from the combined fit to all event categories. 
}
\label{fig:ggmass}
\vspace*{-6mm}
\end{figure}

The Higgs boson mass peak in the diphoton invariant mass distribution 
is clearly visible combining all production modes 
(Fig.~\ref{fig:gg2modes} from~\cite{gammagamma}).
Figure~\ref{fig:ggmass} (from~\cite{gammagamma}) illustrates that in both the ttH 
associated leptonic and hadronic production modes, the current statistics is too 
small to observe a peak in the diphoton mass spectrum.
Table~\ref{tab:gammagamma} (from~\cite{summary}) lists signal, background and data events.
The signal strength
is shown in 
Fig.~\ref{fig:summary} (from~\cite{summary})
for ttH ($\rm H\rightarrow\gamma\gamma$) together with the results from the 
other Higgs boson decay modes.

\begin{table}[htp]
\vspace*{-0.2cm}
\caption{
Expected signal yields (S) and background yields (B) 
in the ttH ($\rm H\rightarrow \gamma\gamma$) search.
The tHqb+WtH yields 
are also presented,
and are included in the background yields.
The effect of background is illustrated by providing the number of background events 
after the fit to data in the smallest diphoton mass interval expected to contain 
90\% of the SM signal events. This corresponds to a diphoton mass interval of [121.9-127.8]
GeV and [121.9-127.9] GeV in the leptonic and hadronic channels, respectively.
\label{tab:gammagamma} }
\begin{center} 
\vspace*{-0.1cm}
\begin{tabular}{cccccc} \hline
Category      & ttH (S) & Background (B)  & tHqb+WtH & S/B & Data \\ \hline
all-hadronic  & 1.58    & 8.27            & 0.10     & 0.19& 9    \\ 
leptonic      & 1.16    & 2.42            & 0.10     & 0.48& 2    \\ \hline
\end{tabular}
\end{center}
\end{table}

\section{ttH Initial Run-2 Results and Comparison with Run-1 Results}

\begin{figure}[h!]
\vspace*{-0.4cm}
\includegraphics[width=0.49\textwidth,height=5cm]{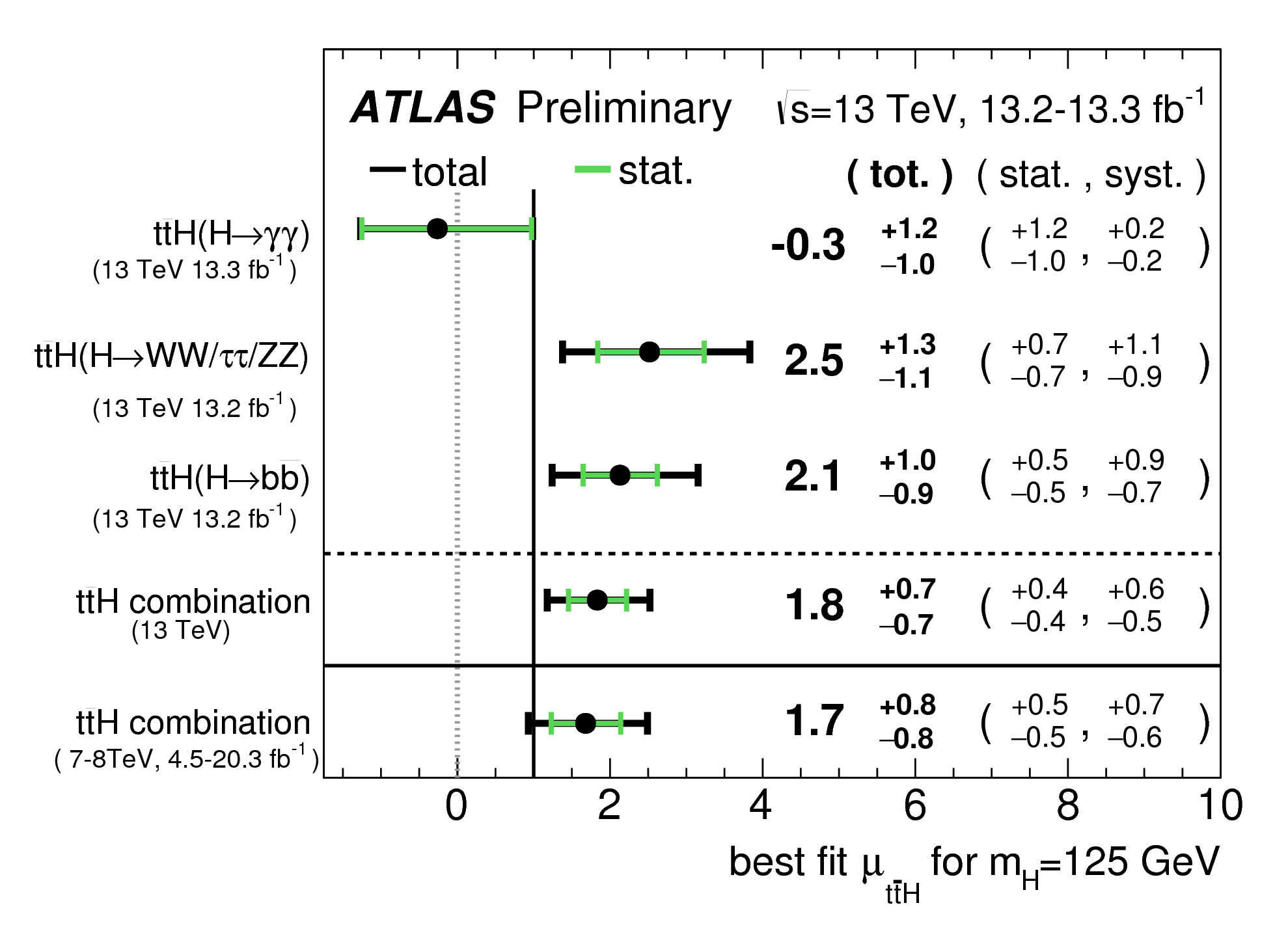} \hfill
\includegraphics[width=0.49\textwidth,height=5cm]{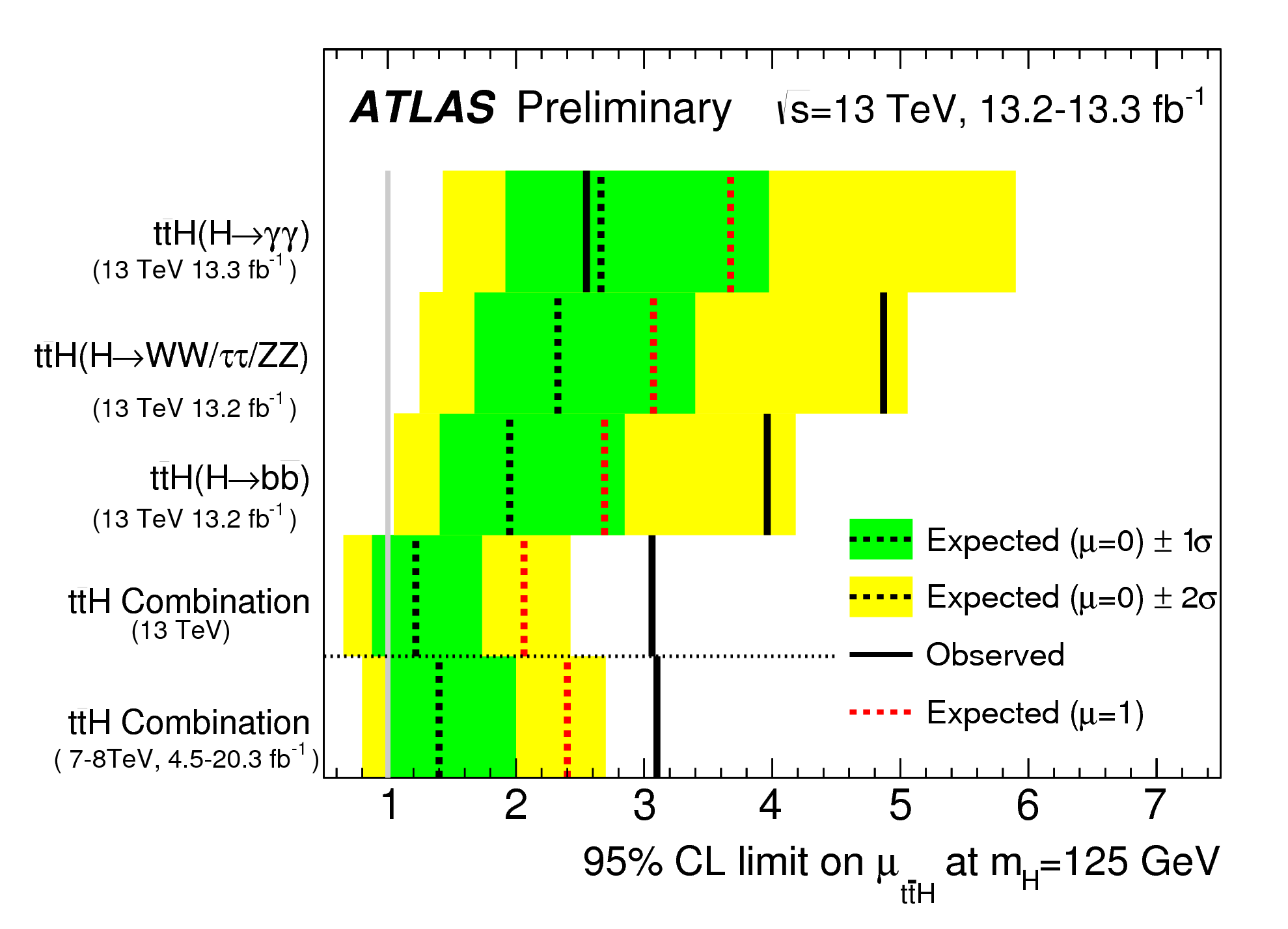}
\vspace*{-0.5cm}
\caption{
Left:
summary of the observed μttH signal strength measurements from the individual analyses 
and for their combination, assuming $m_{\rm H} = 125$~GeV. 
The total (tot.), statistical (stat.), and systematic (syst.) uncertainties on $\mu_{\rm ttH}$ 
are shown. The SM $\mu_{\rm ttH}=1$ (0) expectation is indicated as the black (gray) 
vertical line. 
The observed $\mu_{\rm ttH}$ signal strength measurement obtained from the Run-1 
combination is also presented for comparison (bottom).
Right: 
upper limits on the ttH signal strength for the individual analyses as well as their 
combination at 95\% CL. The observed limits (solid lines) are compared to the expected 
(median) limits under the background-only hypothesis (black dashed lines) and under the 
signal-plus-background hypothesis assuming the SM prediction for the ttH process 
(red dashed lines). The surrounding shaded bands correspond to the $\pm1\sigma$ and 
$\pm2\sigma$ ranges around the expected limits under the background-only hypothesis. 
The vertical grey line at $\mu_{\rm ttH}=1$ represents the point below which the SM 
ttH production would be excluded. The observed and expected limits obtained from the 
Run-1 combination are also shown for comparison (bottom).
}
\label{fig:summary}
\vspace*{-0.5cm}
\end{figure}

The results of the ttH (multilepton) and ttH ($\rm H\rightarrow \gamma\gamma$) 
searches are shown in 
Fig.~\ref{fig:summary} (from~\cite{summary}) together with the results 
from the ($\rm H\rightarrow bb$) search~\cite{tthbb}. 
The uncertainties in the ttH (multilepton) search are dominated by the systematic
uncertainties arising from non-prompt background 
(heavy-flavor),
mis-reconstruction of electric charge, and
irreducible background 
(diboson, ttW, ttZ simulation).
With the increasing statistics from the growing dataset,
these uncertainties will be reduced, together with the 
statistical uncertainties.
In the ttH ($\rm H\rightarrow \gamma\gamma$) search, an analytic function is used to 
fit the data sidebands in order to estimate the continuum background. The resulting systematic 
uncertainty is much smaller than the statistical uncertainty, 
and the increase of analysed data
will directly lead to a significant reduction of the total uncertainty.

The ATLAS measurement of $\mu_{\rm ttH}$ with the initial LHC Run-2 data agrees very well
with the result obtained with the complete LHC Run-1 dataset, both for the mean value
and its upper limit, as summarized in Fig.~\ref{fig:summary} (from~\cite{summary}).

\clearpage
\section{Conclusions}
The direct ttH measurement is the key to determine the top Yukawa coupling 
independent of the physics in the gluon-Higgs sector.
Several analysis channels contribute sensitivity.
The excellent initial (2015 and 2016) LHC Run-2 operation and the 
efficient ATLAS detector operation allowed to record about 39~fb$^{-1}$ data of 
which about 13~fb$^{-1}$ are analysed up to now.
The ttH multilepton and diphoton analyses 
do already have a higher sensitivity with initial Run-2 data if compared to the complete 
Run-1 results.
There is no significant deviation from the SM expectation. 
The initial Run-2 results are very well in agreement with the Run-1 results, both
having a mild data excess.
The prospects of increasing the $\mu_{\rm ttH}$ measurement precision with the full Run-2 dataset
are excellent. 
This could lead to 
strong exclusions of models predicting non-SM ttH rates or an indication of New Physics.

\vspace*{-0.3cm}
\section*{Acknowledgments}
I would like the thank the colleagues from the ATLAS and CMS Higgs working groups, and
the theorists and phenomenologists present at the conference for
the fruitful discussions, as well as
the organizers of HC2016 for their invitation and hospitality.
The project is supported by the 
Ministry of Education, Youth and Sports of the Czech Republic under projects number 
LG 15052 and LM 2015058.

\end{document}